\documentclass[10pt,journal]{IEEEtran}

\usepackage{amssymb}
\usepackage{amsfonts}
\usepackage{graphicx}
\usepackage{amsmath}
\usepackage{subfigure}
\usepackage{array}
\usepackage{color}
\newcommand{\remove}[1]{}

\remove{\evensidemargin 0mm \oddsidemargin 0mm \textwidth 150mm
\textheight 242mm \topmargin -0.7in
}

\newtheorem{theorem}{Theorem}
\newtheorem{corollary}{Corollary}
\newtheorem{lemma}{Lemma}

\newtheorem{definition}{Definition}

\newcommand{\Ua}[0]{\makebox[1ex][l]{\lower.15ex
                                 \hbox{$\uparrow$}}\kern-1ex\lower-.15ex
                                 \hbox{$\uparrow$}}

\newcommand{\Da}[0]{\makebox[1ex][l]{\lower.15ex
                                 \hbox{$\downarrow$}}\kern-1ex\lower-.15ex
                                 \hbox{$\downarrow$}}

\newcommand{\bm}{\bibitem}

\newcommand{\nin}[0]{\makebox[1ex][l]{\lower.15ex
                                \hbox{$\in$}}\kern-.7ex
                                \hbox{$/$}}
\newcommand{\neqn}[0]{\makebox[1ex][l]{\lower.15ex
                                \hbox{$\leq$}}\kern-.7ex
                                \hbox{$/$}}

\makeatletter

\newcommand{\Rmnum}[1]{\expandafter\@slowromancap\romannumeral #1@}
\makeatother

\begin{document}
\bibliographystyle{unsrt}

\title{\bf\Large  Message Authentication Code over a Wiretap Channel}

\author{Dajiang Chen, \and  Shaoquan Jiang,  \and Zhiguang Qin
 \thanks{Dajiang Chen and Zhiguang Qin are  with
 School of Computer Science and Engineering,
  University of Electronic Science
and Technology of China,  Chengdu 611731, China.
\textsf{Email:\quad dajiangchen2010@gmail.com.}
}
\thanks{Shaoquan Jiang is with Institute of Information Security,  Mianyang Normal University, Mianyang 621000, China. \textsf{Email:\quad shaoquan.jiang@gmail.com}}
}
\maketitle

\begin{abstract}
Message Authentication Code (MAC) is a keyed function $f_K$ such that when Alice, who shares the  secret $K$ with Bob,  sends  $f_K(M)$  to  the latter, Bob will be assured of the integrity and authenticity of $M$. Traditionally,  it is assumed that  the channel is noiseless.  However,  Maurer showed that in this case  an attacker  can succeed with  probability $2^{-\frac{H(K)}{\ell+1}}$ after authenticating  $\ell$ messages. In this paper, we consider the setting where the  channel is   noisy. Specifically, Alice and Bob are connected by a discrete memoryless channel (DMC) $W_1$ and a noiseless but insecure channel. In addition, an attacker Oscar is connected with Alice through DMC $W_2$ and with Bob through a noiseless channel.  In this setting, we study the framework that sends $M$ over the noiseless channel and the traditional MAC $f_K(M)$ over channel  $(W_1, W_2)$. We regard the noisy channel as an expensive resource and define the authentication rate $\rho_{auth}$ as the ratio of message length to the number $n$ of channel $W_1$ uses.  The security of this framework depends  on the channel coding scheme for $f_K(M)$. A natural coding scheme is to use the secrecy capacity achieving code of Csisz\'{a}r and K\"{o}rner. Intuitively, this is also the optimal strategy. However, we propose a coding scheme that achieves a higher $\rho_{auth}.$  Our crucial point for this is that in the secrecy capacity setting, Bob needs to recover $f_K(M)$ while in our coding scheme this is not necessary. How to detect the attack without recovering $f_K(M)$ is the main contribution  of this work. We achieve this through random coding techniques.
\end{abstract}

\section{Introduction}
In cryptography, a Message Authentication Code (MAC) is a short piece of information
 used to authenticate a message that it was sent by a specified legitimate sender and to
 provide integrity assurance on the message.
Toward this, we must first specify an  adversary model. That is, what an attacker can do and how much power he has.  A widely adopted  model is to  allow an attacker to play a man-in-the-middle attack. Under this, an attacker Oscar can send any message  to  receiver Bob in the name of legitimate sender Alice.  Besides, any message from Alice must first go through Oscar, who can choose to block, modify or faithfully deliver it. Finally, Oscar is said to  {\em succeed} if Bob accepts a source message $M$ while Alice has never authenticated it. To prevent attacks, Alice and Bob usually share  a secret key $K.$ If the attacker tries to authenticate a source message to Bob before seeing any communication between Alice and Bob, it is called an {\em impersonation attack.} If the attacker tries to modify the message from Alice so that Bob accepts it as an authentication of another source message, it  is called a {\em substitution attack.} In this paper, we study the above general model where the attacker can play an arbitrary man-in-the-middle attack and see a polynomial number of message authentications.

An adversary power is usually defined in two classes: computationally bounded or  unbounded. In the first class,  an adversary only has a polynomial computing power. In the second class, an adversary has  an infinite computing power. In this paper, we are interested in an unbounded  adversary. In our work, the attacker Oscar attempts to fool Bob to accept a fake authentication. Since a legal Bob is always polynomially bounded, we will restricted Oscar to activate Bob with incoming messages for a polynomial number of  times.

Usually, message authentication implicitly assumes the communication channel between Alice and Bob is noiseless. For a detailed treatment, see Simmons \cite{Simmons-Auth} and also Maurer \cite{Maurer}.
However, under this model, any new authentication will cause  an  entropy loss of the secret key and the adversary success probability will increase. In fact, Maurer \cite{Maurer} showed that after  $\ell$ times of authentication, an adversary can succeed in an attack with probability at least $2^{-H(K)/(\ell+1)},$ which quickly approaches 1 with $\ell$. In this paper, we investigate the authentication problem where the channel is noisy.

\subsection{Related works}
A noise in the real world usually {plays} an unwanted role. The task of digital communication is mainly to remove the effect of a channel noise.
However, in 1975,  Wyner \cite{Wyner} was trying to guarantee that the rate of leaked information went to zero as block-length goes to infinity.
In his model, the channel from Alice to Bob is less {noisy} than one between Alice and the attacker. This result was generalized by  Csisz\'{a}r and K\"{o}rner  \cite{Csiszar1}. Since then, the secret sharing problem has been extensively studied (e.g., \cite{Maurer1,Maurer03,Maurer2,Renna}).

Even though secret sharing over noisy channels has been extensively studied, the attention to its sibling {\em message authentication} is far from enough.
Korzhik et al \cite{KY+07} considered the authentication problem over a (noiseless) public discussion channel under the initialization from the  noisy channels so that the sender, the receiver and the attacker hold some correlated data.  So they essentially considered the authentication in the noiseless channel with a noisy initialization (or simply in the source model \cite{Csiszar4}). This framework was further studied in \cite{Barni}.
 Lai, ElGamal and Poor \cite{Lai} considered the authentication over a wiretap channel $X\rightarrow (Y, Z)$. When Alice sends $X$, Bob will receive $Y$ via a DMC $W_1: \mathcal{X}\rightarrow \mathcal{Y}$ and the attacker will receive $Z$ via DMC $W_2: \mathcal{X}\rightarrow \mathcal{Z}.$  Alice and Bob share a secret $K$. The channel between the attacker and Bob is noiseless. They showed that as long as $I(X; Y)>I(X; Z)$, they can build an authentication protocol which can authenticate many source messages without significantly increasing an adversary success probability.  From Maurer \cite{Maurer}, this is impossible when the channel is completely noiseless. Baracca, Laurenti and Tomasin \cite{BLT12} studied the authentication problem over  MIMO  fading wiretap channels. They protocol assumes no shared key between Alice and Bob. They only considered  an impersonation attack and also assume  an authenticated way for a sender to send some preliminary data to a receiver.  This framework was further studied in \cite{FL+13}.

\subsection{Contribution}

\begin{figure}[t]
\centering
\includegraphics[scale=0.40]{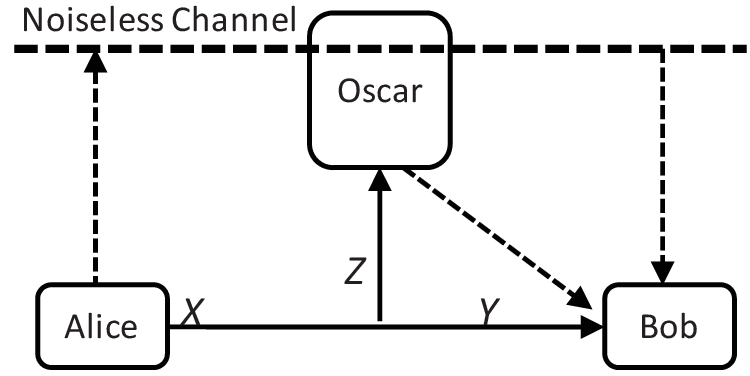}
\caption{The communication model.}\label{Fig.1}
\end{figure}

In this paper, we consider an authentication model as follows. A legitimate transmitter Alice plans to send a message and authenticate it to receiver Bob in the presence of an adversary Oscar. It is assume that Alice and Bob share a secret key $K$, and there is  a DMC $W_{1}:\mathcal{X}\rightarrow {\mathcal{Y}}$ from Alice to Bob and an one-way noiseless channel from Alice to Bob goes through Oscar.
In addition, there is a DMC  $W_{2}:\mathcal{X}\rightarrow {\mathcal{Z}}$ from Alice to Oscar and a noiseless channel from Oscar to Bob {(see Fig. \ref{Fig.1} for an illustration)}.
Practically, Internet, Telephone, or a wireless communication system with an error correcting-code can serve as this channel.
We also assume that Oscar has unbounded computing resources, and can play an arbitrary man-in-the-middle attack  and see a polynomial number of message authentications (details in Section \ref{Model}).

\begin{figure}[t]
\centering
\includegraphics[scale=0.40]{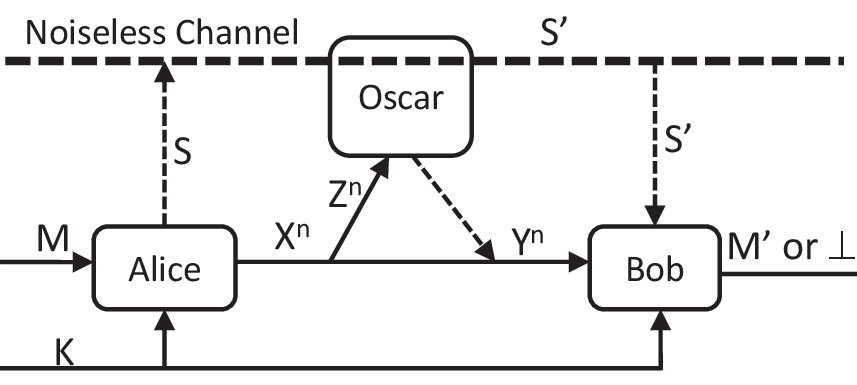}
\caption{The authentication model.}\label{Fig.2}
\end{figure}

We study the message authentication code (MAC) in these noisy model {(see Fig. \ref{Fig.2} for an illustration)}: Alice encodes $M$ to a codeword  $(S, X^n)$ with $S$ sent over the noiseless channel and $X^n$ sent over the wiretap channel $(W_1, W_2)$, which arrives at Bob as $(S',Y^n)$, where $n$ is the length of codeword over the  noisy  channel $W_1$ and   $S'
$ is the received version of $S$ by Bob through Oscar.
Upon $(S', Y^n)$, Bob decides to reject or accept the authentication by checking the consistency of $(S', Y^n).$
We regard the transmission  over the wiretap channel as an expensive resource and define an efficiency measure for the MAC as {\em authentication rate} $\rho_{auth}=\frac{|M|}{n}$, where  $|M|$ denotes  the bit length of $M$.

The goal of the paper is to propose a  MAC protocol in the above model achieving a multi-messages authentication  with the same secret key $K$.
The main efficiency criterion is a minimization of the wiretap channel usage while keeping the probability of Oscar mounting a successful attack
negligible.
To achieve the goal with high efficiency, we present a natural MAC scheme as follows.
Alice first generates a traditional message authentication code  $T$ of $M$ and uses an channel coding to encode $T$ to $X^n$.
Finally, the codeword  is $(M, X^n)$, where $M$ is for  the noiseless channel and $X^n$ is for channel $(W_1, W_2)$.
Upon $(M', Y^n)$, Bob's verification is to check the consistency of $M'$ and $Y^n$.
The main challenge is how to design a channel coding with completeness and authentication (details of completeness and authentication in Section \ref{Model}). We addresses this issue by leveraging random coding techniques (details in Section \ref{Scheme}).

In the proposed scheme, we can rewrite  $\rho_{auth}=\rho_{tag}\cdot\rho_{chan}$,
 where $\rho_{chan}=\frac{|T|}{n}$ is called {\em channel coding rate} and  $\rho_{tag}=\frac{|M|}{|T|}$ is called {\em  the rate of tag}.
$\rho_{tag}$  is mainly determined by purely cryptographic techniques while $\rho_{chan}$ is determined by channel coding techniques.
The latter is our main focus.
With secrecy capacity $C_s$ of Csisz\'{a}r and K\"{o}rner \cite{Csiszar1} in mind, if we naturally encode $T$ to $X^n$ using their code, we get $\rho_{auth}=\rho_{tag} C_s$, which can be trivially generalized from Lai's work \cite{Lai}.  This  intuitively seems to be the best possible result as we have to protect $T$ in its full secrecy. However, we propose a new coding technique, achieving $\rho_{auth}=\rho_{tag} (H(X|Z)-\delta)$ for any small $\delta>0.$  As shown in \cite{Csiszar1}, when  channel $W_1$ is less noisy than channel $W_2$, then $C_s=H(X|Z)-H(X|Y)$. So the ratio of authentication rate of the natural scheme to ours is arbitrarily close to $1-H(X|Y)/H(X|Z)<1$.  Our crucial point for this is that the secrecy capacity guarantees that Bob can recover $T$ while in the setting, this is unnecessary because in case of no attack, it can be computed from $M'=M$ and $K$ and in case of an attack, Bob only needs to detect the inconsistency of $M'$ and $Y^n$ and reject.  How to detect the attack  without fully recovering  $T$ is the non-trivial part of our work.  We achieve this through random coding techniques.\\

This paper is organized as follows. Section II introduces  basic concepts and results that will be used  in this paper. Section III introduces our authentication model. Section IV introduces our MAC. Section V proves an  authentication theorem of our MAC. Section VI discusses the efficiency of our MAC. The last section is a conclusion.
\section{Preliminaries}

In this paper, we use the following notations or conventions.
\begin{itemize}
\item A random variable (RV) is  denoted by an upper case letter (e.g., $X, Y$); its realization is  denoted by a lower case letter (e.g., $x, y$);  its  domain is denoted by a calligraphic letter (e.g.,
$\mathcal{X}, \mathcal{Y}$); $X\leftarrow_U\mathcal{X}$ means that $X$ is chosen uniformly at random from $\mathcal{X}$.
\item $x^n$ denotes a sequence $x_1, \cdots, x_n$  of length $n$. For a positive integer $s$, define $[s]=\{1, \cdots, s\}.$
\item Probabilities $P(X=x)$ and $P(X=x|Y=y)$ are denoted by $P_X(x)$ and $P_{X|Y}(x|y)$. $P^n_{Y|X}(y^n|x^n)$ is defined as $\prod_{i=1}^nP_{Y|X}(y_i|x_i) $
\item $P_{x^n}(\cdot)$ is  a distribution over alphabet $\mathcal{X}$, where   for any  $a\in \mathcal{X}$, $P_{x^n}(a)$ is the fraction of $a$ in sequence $x^n$. Similarly, $P_{x^ny^n}(a, b)$ is the fraction of $(a, b)$ in sequence $(x_1, y_1), (x_2, y_2), \cdots, (x_n, y_n)$.
\item \emph{Distance} between  RVs $X$ and $X'$ is  $\textsf{SD}(X; X')=\sum_{x} |P_X(x)-P_{X'}(x)|.$ \emph{Conditional
     distance between $X$ given $Y$} and  $X'$  is
    \begin{eqnarray*}
     \textsf{SD}(X|Y; X') = \sum_{y, x} P_Y(y) |P_{X|Y}(x|y)-P_{X'}(x)|.
      \end{eqnarray*}
\item Entropy
       $H(X)=-\sum_{x}P_X(x)\log P_X(x)$; mutual information $I(X; Y)= \sum_{x, y}P_{XY}(x, y)\log \frac{P_{XY}(x, y)}{P_X(x)P_Y(y)}$;    conditional entropy
      \begin{eqnarray*}
      H(X|Y)=-\sum_{x, y}P_{XY}(x, y)\log P_{X|Y}(x|y).
      \end{eqnarray*}
\item RVs $X_1, \cdots, X_m$ form a {{Markov chain}},
      denoted by $X_1\rightarrow\cdots\rightarrow X_m$,  if $P_{X_i|X_{i-1}\cdots X_{1}}(x_i|x_{i-1}\cdots x_{1}) = P_{X_i|X_{i-1}}(x_i|x_{i-1}),$ for any $i=2, \cdots, m.$
\end{itemize}

The following lemma \cite[Lemma 1]{Csiszar2} shows the relationship between mutual information and distance.
\begin{lemma}\label{lem1} {\cite{Csiszar2}} Let $X$ and $Y$ be two RVs over ${\mathcal X}$ and ${\mathcal Y}$ respectively. $|{\mathcal X}|\ge 4.$ Let $\Delta=\textsf{SD}(X|Y; X).$ Then
\begin{equation}
\frac{1}{2\ln 2}\Delta^2\le I(X; Y)\le \Delta\log \frac{|{\mathcal X}|}{\Delta}.  \label{le: Ibound}
\end{equation}
\end{lemma}

\subsection{$\epsilon$-almost strongly universal hash function}

A universal hash function essentially is a family of compression functions that has an almost uniformly distributed  output.
It was introduced by Wegman and Carter \cite{WC79} and further developed in \cite{WC81,Stin94}.
We now introduce the    $\epsilon$-almost strongly-universal hash function.
\begin{definition} A finite family of hash functions $H$ from alphabet $\mathcal{M}$ to a finite alphabet $\mathcal{T}$ is \emph{$\epsilon$-almost strongly-universal} ($\epsilon$-ASU) if the following holds
\begin{itemize}
\item[-]  $|\{h\in \mathcal H: h(x)=t\}|=\frac{|\mathcal H|}{|\mathcal{T}|}$, $\forall x\in \mathcal{M}, \forall t\in \mathcal{T}$,

\item[-]
$|\{h\in \mathcal H: h(x_1)=t_1, h(x_2)=t_2\}|\le  \frac{\epsilon |\mathcal H|}{|\mathcal{T}|}, \forall x_1, x_2\in \mathcal{M}$ ($x_1\ne x_2), \forall t_1, t_2\in \mathcal{T}.$
\end{itemize}
\end{definition}
We remark that domain $\mathcal{M}$ is not necessarily finite but $\mathcal{T}$ and $\mathcal H$ are both finite. In this paper, $\mathcal H$ is indexed by elements in a set $\mathcal{K}.$  We can  write $\mathcal H=\{h_k\}_{k\in \mathcal{K}}.$ So  $|\mathcal{K}|=|\mathcal H|$ and $h_k$ is uniformly random in $\mathcal H$ when $k$ is so in $\mathcal{K}.$

A construction of  $\epsilon$-ASU hashing function with a good input/output  ratio  will be used in this paper. Stinson \cite{Stin94} showed that there exists a scheme that compresses $2^s\log q$ length to $\log q$ length. We state it as follows.
\begin{lemma}\cite{Stin94}
Let $q$ be a prime power and let $s\geq 1$
be an integer. Then there exists an $\frac{s}{q}$-ASU hash function
 from $\mathcal{M}$ to $\mathcal{T}$ with key space $\mathcal{K}_0$, where  $|\mathcal{K}_0|=q^{s}$, $|\mathcal{M}| = q^{2^s}$, $|\mathcal{T}|= q$. \label{le: stin}
\end{lemma}

\remove{Note that if we take $s=\log q=\ell$, then in this hash family the output has length $\ell$ bits, the input has length $\ell 2^\ell$ bits and it can be indexed by $K$ of length $\ell^2$ bits. Later our input represents the message to be authenticated. So if we use this family of hash function, this means we can use a key of length  $\ell^2$ to authenticate a message of an exponential length (i.e., $\ell 2^\ell$) such that the tag has length only $\ell.$ }

\subsection{Discrete memoryless channel}
A {\em discrete memoryless channel} (DMC) with input alphabet $\mathcal{X}$ and output alphabet $\mathcal{Y}$ is characterized by  a
stochastic matrix  $W=\{W(y|x)\}_{x\in \mathcal{X}, y\in \mathcal{Y}}$, where $W(\cdot |x)$ is the distribution of the channel output $Y$ when  the input is  $X=x$, i.e.,
$W(y|x)=P_{Y|X}(y|x)$. In this case, we say $X$ and $Y$ are {connected by channel} $W$. If the input sequence is $x^n$ and the output sequence is $y^n$, then $P_{Y^n|X^n}(y^n|x^n)=
\prod_{i=1}^n P_{Y|X}(y_i|x_i)=\prod_{i=1}^n W(y_i|x_i).$    For simplicity, we  denote $\prod_{i=1}^n W(y_i|x_i)$ by $W(y^n|x^n)$.

{\em  A $n$-length code} ${\cal C}$ for a DMC $W: \mathcal{X}\rightarrow \mathcal{Y}$ with message space $\mathcal{T}$ is a pair of functions $(f, \phi)$, where $f:  \mathcal{T}\rightarrow \mathcal{X}^n$ is the encoding function,  $\phi: \mathcal{Y}^n\rightarrow \mathcal{T}\cup\{\perp\}$ is the decoding function, and  $\perp$ denotes a detection of  error. For $t\in \mathcal{T}$, $f(t)\in \mathcal{X}^n$ is called a {\em codeword}. When a sender wants to send message $t$, he sends $f(t)$. When a receiver receives vector $y^n\in \mathcal{Y}^n$, he decodes it to $\phi(y^n).$ If $\phi(y^n)\ne t,$ an error occurs.    The {\em error probability} of a code
is defined $e({\cal C}) =P(\phi(Y^n)\ne T)$, where $Y^n$ is the channel output with  message $T$ that  is uniformly random over $\mathcal{T}$.

\subsection{Typical sequences}

Let $x^n$ be a sequence over $\mathcal{X}$. Then the distribution $P_{x^n}(\cdot)$ is called the {\em type}  of $x^n$ over $\mathcal{X}$, where $P_{x^n}(a)$ is the fraction of  occurrences of $a$ in $x^n$.   For a type $P$ over $\mathcal{X}$, {\em type set} $\textsf{T}_{P}^n$ denotes  the set of all $n$-length sequences over $\mathcal{X}$ with type $P.$



\begin{definition} Let $X$ be a RV over alphabet $\mathcal{X}$.  $x^n\in \mathcal{X}^n$ is    {\em $\epsilon$-typical} if $|P_{x^n}(a)-P_X(a)|\le \frac{\epsilon}{|\mathcal{X}|}$ for all $a\in \mathcal{X}$, and further it holds that  $P_{x^n}(a)=0$ for any $a$ with  $P_X(a)=0$.  The set of $\epsilon$-typical sequences for $X$ is denoted by $\textsf{T}_{[X]_\epsilon}^n$.
\end{definition}

Note that if $x^n\in \textsf{T}_{[X]_\epsilon}^n$, then the whole type set $\textsf{T}_{P_{x^n}}^n$ is included in $\textsf{T}_{[X]_\epsilon}^n.$ So  $\textsf{T}_{[X]_\epsilon}^n$ is a union of some type sets whose type is  ``close'' to $P_X.$
 Note the form of  $X$ could be arbitrary. Especially, it  could be a vector such as $X=(Y,Z).$ If $x^n=(y^n, z^n)$ is $\epsilon$-typical for $X=(Y, Z)$, we say $(y^n, z^n)$ is {\em jointly $\epsilon$-typical}. The set of jointly $\epsilon$-typical sequences for $Y$ and $Z$ is denoted by $\textsf{T}_{[YZ]_\epsilon}^n.$

\begin{definition} Let $X$ and $Y$ be RVs over alphabet $\mathcal{X}$ and $\mathcal{Y}$ respectively. $y^n\in \mathcal{Y}^n$ is {\em conditionally $\epsilon$-typical} given    $x^n\in \mathcal{X}^n$,  if $|P_{x^ny^n}(a, b)-P_{x^n}(a)P_{Y|X}(b|a)|\le \frac{\epsilon}{|\mathcal{X}|\cdot|\mathcal{Y}|}$ for all $a\in \mathcal{X}$ and $b\in \mathcal{Y}$, and further it holds that  $P_{x^ny^n}(a, b)=0$ for any $a, b$ with  $P_{XY}(a, b)=0$.
The set of conditionally $\epsilon$-typical sequences for $Y$, given $x^n$, is denoted $\textsf{T}_{[Y|X]_\epsilon}^n(x^n).$ If $X$ and $Y$ are connected by DMC $W$,  $\textsf{T}_{[Y|X]_\epsilon}^n(x^n)$ is also denoted by $\textsf{T}_{[W]_\epsilon}^n(x^n).$
\end{definition}

We now introduce some basic properties of  typical sequences, which are well-known and can be found in
existing information theory books (e.g. \cite[Chap 1.2]{CC2011}).
\begin{lemma} Let $X_1, X_2, X$ be RVs over $\mathcal{X}$ and $Y$ be  a RV over $\mathcal{Y}$. Then,

\begin{itemize}

\item[1.] For  any type $Q$ of $\mathcal{X}^n$,
\begin{equation*}
(n+1)^{-|\mathcal{X}|}\cdot 2^{nH(Q)}\le |\textsf{T}_Q^n|\le 2^{nH(Q)}.
\end{equation*}
\item[2.]  There exists constant $c>0$ s.t.  for  $\forall \epsilon>0$,  $\forall x^n\in \textsf{T}_{[X]_\epsilon}^n$,
\begin{eqnarray*}
\nonumber
2^{-n[H(X)+c\epsilon]}\le P_X^n(x^n)\le 2^{-n[H(X)-c\epsilon]},\\
\nonumber
(1-\epsilon)2^{n[H(X)-c\epsilon]}\stackrel{*}{\le} |\textsf{T}_{[X]_\epsilon}^n|\le 2^{n[H(X)+c\epsilon]}.
\end{eqnarray*}
where inequality ($*$) holds when $n$ large enough.
\item[3.] There exists constant $c>0$ s.t.  for  $\forall \epsilon>0$, $\forall x^n\in \textsf{T}_{[X]_\epsilon}^n$, $\forall   y^n\in \textsf{T}_{[Y|X]_\epsilon}(x^n)$,
\begin{eqnarray}
\nonumber
   2^{-n[H(Y|X)+c\epsilon]}\le P_{Y|X}^n(y^n|x^n)\le 2^{-n[H(Y|X)-c\epsilon]}, \\
\nonumber
   (1-\epsilon)2^{n[H(Y|X)-c\epsilon]}\stackrel{*}{\le}|\textsf{T}_{[Y|X]_\epsilon}^n(x^n)|\le 2^{n[H(Y|X)+c\epsilon]}.
\end{eqnarray}
where inequality (*)  holds when $n$ large enough.

\item[4.]  There exists  constants $\lambda_1$ and $\lambda_2>0$ such that when $n$ large enough, for any $x^n\in \textsf{T}_{[X]_\epsilon}^n$
\begin{eqnarray*}
 P_Y^n(\textsf{T}^n_{[Y]_\epsilon})&\ge& 1-2^{-n\lambda_1\epsilon^2},\\
 P_{Y|X}^n(\textsf{T}^n_{[Y|X]_\epsilon}(x^n)|x^n)&\ge& 1-2^{-n\lambda_2\epsilon^2}.
\end{eqnarray*}

\end{itemize} \label{le: basicH}
\end{lemma}


\section{MAC for  a wiretap channel: the model}\label{Model}

\textbf{{Syntax Model. }}\quad
Assume that there is  a DMC $W_{1}:\mathcal{X}\rightarrow {\mathcal{Y}}$ from Alice to Bob.
There is also an one-way noiseless channel from Alice to Bob.
There is a DMC  $W_{2}:\mathcal{X}\rightarrow {\mathcal{Z}}$ from Alice to Oscar and a noiseless channel from Oscar to Bob.
In this section, we will formulate the  {\em message  authentication code} in  this channel model. It  allows Alice to authenticate a message to Bob while preventing attacks from Oscar.  Let ${\mathcal M}$ be the message space.  The system is described by an {\em encoding function}  $F:\mathcal M\times \mathcal K\rightarrow  \mathcal{S}\times{\mathcal{X}^n}$ and a {\em decoding function} $G: \mathcal{K}\times\mathcal{S}\times \mathcal{Y}^n\rightarrow \mathcal{M}\cup\{\bot\}$.  The authentication syntax is as follows.
\begin{itemize}
\item If Alice wishes to authenticate $M\in \mathcal{M}$ to Bob, she  computes  $(S, X^n)=F(M, K).$
She then  sends  $S$ over  a noiseless channel to Bob, and sends $X^n$ over a wiretap channel $(W_1, W_2)$.  Through Oscar,  $S$ will arrive at Bob as $S'$. Let $X^n$, through $W_1$, arrive at Bob as $Y^n$ and, through $W_2$, arrive at Oscar as $Z^n.$
 \item Upon $(S', Y^n)$, Bob  computes  $M'=G(K, S', Y^n)$. If $M'\ne \perp$, he outputs $M'$ as the authenticated message from Alice; otherwise, he rejects.
\end{itemize}

Note if Alice does not use the noisy channel, then our model degenerates to a traditional MAC model.  So naturally, we call $(F, G)$ \emph{a message authentication code (MAC) over  channel $(W_1,W_2)$} and call $(S, X^n)$ the {\em codeword} of $M$.
For our convenience, we define a \emph{decision bit} $D: \mathcal{K}\times\mathcal{S}\times \mathcal{Y}^n\rightarrow \{0, 1\}$ such that $D(K,S', Y^n)=0$ if and only if $G(K,S', Y^n)=\bot$ (i.e., Bob rejects).

\vspace{0.05in} \noindent {\bf Adversary Model. } \quad
An authentication failure could come from a completeness error  or  an  attack from Oscar.
If MAC is designed properly, the completeness error is  negligible.
So we focus on attacks.  In our model, Oscar can arbitrarily  modify  $S$ over the noiseless channel.
We assume that the channel from Oscar to Bob is noiseless and hence Oscar can launch impersonation attacks.
We also allow Oscar to learn the decision bit for each authentication. Granting  Oscar to learn this is not artificial.
For instance, if Bob rejects the authentication, he could request Alice to re-authenticate the message.
This allows  Oscar to learn the decision  bit $b=0$.
For another instance, if $M$ is one message in a serial authentication procedure (such as a stream authentication),
  Bob could feedback an  updated message index that represents the current  successfully authenticated message.
This implicitly allows Oscar to learn the decision bit.
We also wish to capture the concern  that even if Oscar has {\em adaptively} attacked many authentication instances,
 he still cannot cheat Bob to accept a false authentication.
The formal model is as follows.
\begin{itemize}
\item[I.] Let $M_i, i=1, 2, \cdots,$ be the sequence of messages authenticated by Alice. Let $(S_i, X_i^n)$ be the codeword  of $M_i$. Alice sends $S_i$ over the noiseless channel to Bob and $X^n_i$ over channel $(W_1, W_2)$.  Oscar can revise $S_i$ to arbitrary $S_i'\in \mathcal{S}.$ Let $X_i^n$ arrive at Bob as $Y^n_i$ and at Oscar as $Z_i^n$. Let $M_i'=G(K, S_i', Y^n_i)$ and $b_i=D(K, S_i', Y^n_i)$.  Besides $(S_i, Z_i^n)$, Oscar also learns $b_i.$

     Note here we consider an adaptive Oscar. So he  determines $S_i'$ based on  $S_i$, his local random source $R$ and the information collected previously:   $\{(S_j, Z^n_j)\}_{j=1}^{i-1}$  and  decision bits $\{b_j\}_{j=1}^{i-1}$ in stage I and decision bits $\{\hat{b}_j\}$ in stage II below.

\item[II.] Oscar can adaptively send $\hat{S}_t\in{\mathcal{S}}$ and $\hat{Y}^n_t\in {\mathcal Y}^n$  to Bob noiselessly. Oscar will learn Bob's decision bit   $\hat{b}_t=D(K, \hat{S}_t, \hat{Y}^n)$.   He succeeds  if $\hat{b}_t=1.$

    Here  $(\hat{S}_t, \hat{Y}^n_t)$ is computed based on $R$ and the information collected previously: $\{(S_j, Z^n_j, b_j)\}$ and $\{\hat{b}_j\}_{j=1}^{t-1}.$
\end{itemize}
 In the model, Oscar can arbitrarily interleave Type I attacks and Type II attacks. We use {\em succ} to denote the success in a Type I or Type II attack.  \\

\vspace{0.05in} \noindent {\bf Authentication Property. } \quad After introducing  the adversary model, we now define the authentication property formally. It consists of completeness and authentication. Completeness essentially states that when Oscar does not present, Bob should accept  $M$ with high probability. Authentication states that Oscar can succeed in the above two types of attacks  only with a negligible probability.
\begin{definition}  A message authentication code $(F, G)$   over channel $W_1: {\mathcal X}\rightarrow {\mathcal Y}, W_2: {\mathcal X}\rightarrow {\mathcal Z}$ is {\em secure} if the following holds (where $n$ is the number of channel $W_1$ uses).
\begin{itemize}
\item[1.] {\bf Completeness. } \quad If Oscar does not present, then Bob rejects with exponentially (in $n$) small probability.
\item[2.] {\bf Authentication. } If the number of Type II attacks is polynomially bounded (in $n$), $\Pr({succ})$ is negligible.
\end{itemize}
\end{definition}

{\em  Remark. } Restriction on the number of Type II attacks is unavoidable as Oscar  can always choose a message $M$ and impersonate with  every possible $(s, y^n)\in \mathcal{S}\times \mathcal{Y}^n$ to Bob. As $\mathcal{S}$ and $\mathcal{Y}^n$ are finite sets, he can always succeed for some pair $(s,y^n).$  The number of Type II attacks is chosen as be polynomially bounded  as each attack will involve Bob (as a verifier) and it is impractical to require him to be in a complexity class beyond a polynomial. For the same reason, the number of Type I attacks is also polynomially bounded (implicitly), although we do not require this.

\vspace{.05in} \noindent{\bf Efficiency. } \quad  We regard the communication over  a wiretap channel as an expensive resource.
It is desired to minimize the use of it. For convenience of analysis, we define the efficiency measure for a  MAC. It is called {\em authentication rate}, defined as $\rho_{auth}=\frac{\log |\mathcal{M}|}{n}$,
  which is the ratio of the source message length to the codeword length.

\section{Our Scheme}\label{Scheme}

\subsection{Random coding theorem}
To construct our scheme, we will prove the existence of a channel coding scheme that satisfies many strong properties.
It  essentially states that, there exists a set  $\mathcal{C}\subseteq \textsf{T}_P^n$ such that:

 (1) $\mathcal{C}$ can be divided into subsets $\{C_{ij}\}_{i,j}$ (as shown in Fig. \ref{codebook}) such that each column  $C_{1j}\cup\cdots\cup C_{\mathbb{I}j}$ is a  code for channel $W_1$;

 (2) If $I$ is uniform random and $J$ is  an arbitrary RV but independent for I, then  for  $\hat{X}^n$ uniformly random over  $\mathcal{C}_{IJ}$ that is  transmitted over a wiretap  channel $(W_1,W_2$), the output $\hat{Z}^n$ of channel $W_2$  is almost independent of $(I,J)$;

 (3) In item 2, if the output of $W_1$ is $\hat{Y}^n$, then for any adversarially chosen $J'$ (that satisfies certain properties), it is unlikely that $\hat{Y}^n$ can be decoded using $g_{J'}$ into a codeword in ${\cal C}_{IJ'}.$  The formal statement is as follows.

\begin{figure}[!ht]
\centering
\begin{tabular}{|c|c|c|c|c|}
\hline
       $C_{11}$ & $\cdots$ & $C_{1j}$ & $\cdots$ & $C_{1\mathbb{J}}$\\
\hline
       $\vdots$ & $\ddots$ & $\vdots$ & $\ddots$ & $\vdots$ \\
\hline
       $C_{i1}$ & $\cdots$ & $C_{ij}$ & $\cdots$ & $C_{i\mathbb{J}}$ \\
\hline
       $\vdots$ & $\ddots$ & $\vdots$ & $\ddots$ & $\vdots$ \\
\hline
       $C_{\mathbb{I}1}$ & $\cdots$ & $C_{\mathbb{I}j}$ & $\cdots$ & $C_{\mathbb{I}\mathbb{J}}$\\
\hline
\end{tabular}
\caption{The codebook used in our construction.}\label{codebook}
\end{figure}

\begin{theorem}\label{le: code}
Let $X, Y, Z$ be RVs  over $\mathcal{X, Y, Z}$ respectively such  that $P_{Y|X}=W_1, P_{Z|X}=W_2$ for DMCs $W_1, W_2$ and that $P_{X}=P$ for a type $P$ over $\mathcal{X}$ with $P(x)>0, \forall x\in \mathcal{X}$. Assume   {$I(X;Y)>I(X;Z)+\tau$} for some $\tau>0$.  Then, for any integers ${\mathbb I}, {\mathbb J}$  with
 \begin{eqnarray}
0&\le& \frac{1}{n}\log{\mathbb J}<H(X|Y)+\tau,\label{I-ineq}\\
      0&\le& \frac{1}{n}\log{\mathbb I}<  {I(X;Y)-I(X;Z)-\tau},\label{J-ineq}
\end{eqnarray}
 there exists  disjoint subsets ${\mathcal C}_{ij}\subset {\textsf{T}_P^n} $,  $i\in [{\mathbb I}], j\in [{\mathbb J}]$ s.t. when $n$ large enough
\begin{itemize}
  \item [1.] For each $j$, $\mathcal{C}_{\cdot j}\stackrel{def}{=}\bigcup_i{{\mathcal C}_{ij}}$ is a  code $(f_j, g_j)$ for channel $W_1$  that has an  exponentially small average error probability, where $f_j$ encodes a message $m$ to the $m$th codeword in $\mathcal{C}_{\cdot j}$.
  \item [2.] For any RV  $J$ over $[{\mathbb J}]$ and $I$ over $[{\mathbb I}]$ with $P_{IJ}=\frac{P_J}{{\mathbb I}},$ let $\hat{Z}^n$ be  the output of channel $W_2$ with input $\hat{X}^n\leftarrow_U  \mathcal{C}_{IJ}$. Then, $I(I,J;\hat{Z}^n)\leq{2^{-n\beta_2}}$, for some $\beta_2>0$ (not depending  on $P_J).$
\item[3.] For any RV  $J$ over $[{\mathbb J}]$ and $I$ over $[{\mathbb I}]$ with $P_{IJ}=\frac{P_J}{{\mathbb I}},$ let $\hat{Y}^n$ be  the output of channel $W_1$ with input $\hat{X}^n\leftarrow_U  \mathcal{C}_{IJ}$.  Assume RV $J'$ over $[{\mathbb J}]$ with $J'\ne J$ satisfying
\begin{itemize}
\item[a.]  $\textsf{SD}(P_{J'J}; P_{J'J|I})\le \delta_1$;
\item[b.]  $J'\rightarrow IJ\rightarrow \hat{X}^n\rightarrow \hat{Y}^n$ is  a Markov chain;
\item[c.]  $P_{J'J}(j',j)\le{\frac{2^{n^\omega}}{{\mathbb J}({\mathbb J}-1)} +d(j',j)}$ for any $j,j'$ and a function $d(\cdot, \cdot)$ s.t.  $\sum_{j',j}d(j',j)<\delta_2$, where  $\omega$ is a constant in $(0, 1)$.

\end{itemize}
 Then, when $n$ large enough,
   \begin{eqnarray*}
 P\Big{(}g_{J'}(\hat{Y}^n)\in \mathcal{C}_{IJ'}\Big{)}&\le 2^{-{n}^\omega}+\delta_1+\delta_2.
   \end{eqnarray*}
\end{itemize}
\end{theorem}

\begin{IEEEproof}
Please refer to the Appendix.
\end{IEEEproof}

\subsection{Construction}

Now we describe the construction of our MAC.   Let $W_{1}:\mathcal{X}\rightarrow{\mathcal{Y}}$, $W_{2}:\mathcal{X}\rightarrow{\mathcal{Z}}$ be the wiretap channel. Assume {$I(X;Y)>I(X;Z)+\tau$} for some $\tau>0$ and $P_X$ be a type $P$ with  $P(x)>0 $ for any $x\in \mathcal{X}$. Let ${\mathcal C}_{ij}, i=1, \cdots, {\mathbb I}, j=1, \cdots, {\mathbb J}$ be
the subsets of $\textsf{T}_P^n$ obtained in Theorem \ref{le: code}.  Set $\mathcal{K}_1=\{1, \cdots, {\mathbb I}\}.$  Let $h: \mathcal{M}\times \mathcal{K}_0\rightarrow \mathcal{T}$ be a  $\epsilon$-ASU hash function with key space $\mathcal{K}_0$, where range $\mathcal{T}\subset \{1, \cdots, {\mathbb J}\}$.  Alice and Bob share a secret key $K=(K_0,K_1)\in {\mathcal K}_0\times {\mathcal K}_1.$ We now describe the encoding and decoding procedures.
\begin{itemize}
\item[1.] {\em Encode. } To authenticate $M$, Alice computes $T=h_{K_0}(M)$, and randomly takes   ${X}^n$ from ${\mathcal C}_{K_1T}$. Then the codeword of $M$ is $(M, X^n)$, where  $M$ is sent over the noiseless  channel and ${X}^n$ is sent over channel $(W_1, W_2)$.
 \item[2.] {\em Decode. } Upon $(M', {Y}^n)$, Bob computes  $T'=h_{K_0}(M')$. If $g_{T'}(Y^n)\in \mathcal{C}_{K_1T'}$,  he accepts $M'$; otherwise, he rejects, where $g_{j}$ is the decoder of code $\mathcal{C}_{\cdot j}$.

     {\em Note:}  In the code $\mathcal{C}_{\cdot j}$ at Theorem \ref{le: code},  encoder $f_{j}$ encodes message $\ell$ to the $\ell$th codeword in $\mathcal{C}_{\cdot j}$, $g_{j}(Y^n)$ must decode to $\perp$ or a codeword's index  in $\mathcal{C}_{\cdot T'}$. As  an index is 1-1 correspondent to its codeword,  we assume $g_{j}(Y^n)$ decodes to $\perp$ or the   codeword itself.
\end{itemize}

\section{Security Analysis}

In this section, we prove the authentication property  of our MAC. We begin with two  lemmas. The first lemma shows that Oscar obtains no significant amount of information about secret key $(K_0,K_1)$, after {\em eavesdropping} $J$ times of authentications that gives  Oscar information $M_1Z_1^n\cdots M_JZ_J^n$. The idea is as follows. Let $T_j=h_{K_0}(M_J)$. From Theorem \ref{le: code}(2), it is easy to see that $I(K_1T_j; Z_j^n|M_j=m_j)\approx 0.$ Note that given $M_j=m_j$, $K_0K_1\rightarrow K_1T_j\rightarrow Z_j^n$ forms a Markov chain as $Z_j^n$ is decided by $K_1T_j$ and some randomness independent of $K_0K_1$. So by data processing inequality, $I(K_0K_1; Z_j^n|M_j=m_j)\le I(K_1T_j; Z_j^n|M_j=m_j)\approx 0.$
As $K_0K_1$ is independent of $M^J$, $I(K_0K_1; M^JZ^n_1\cdots Z_J^n)=I(K_0K_1; Z_1^n\cdots Z_J^n|M^J).$ Finally, by standard information theory techniques, we can show that  this is bounded by $\sum_{j=1}^J I(K_0K_1; Z_j^n|M^J=m^J)$, which is now known small.  The lemma follows by averaging on $M^J.$ We implement the strategy formally in the following lemma.
\begin{lemma}\label{pro:sec}
Let  $(K_0,K_1)$ be uniformly distributed over $\mathcal{K}_0\times{\mathcal{K}_1}$ and  $M_1, \cdots, M_J$ be $J$ arbitrary messages in ${\cal M}$ authenticated by Alice.
For $j=1, \cdots, J,$ let  $Z^n_j$ be  the output of $W_2$ when Alice sends $X^n_j$ (w.r.t. $M_j$).  Then,  there exists $\beta_2>0$ such that when $n$ large enough,
\begin{equation}
I(K_0K_1;M_1Z^n_1 \cdots M_{J} Z^n_{J})\leq{J\cdot2^{-n\beta_2}}.   \label{eq: I(KZ)J}
\end{equation}
\end{lemma}

\begin{IEEEproof}   For $j=1, \cdots, J,$ let  $T_j=h_{K_0}(M_j). $  Define $M^J=M_1\cdots M_J$ and $m^J=m_1\cdots m_J$ for $m_j\in {\cal M}.$ Then,  $P_{K_1 T_j|M^J}(k_1, t_j|m^J)=P_{K_1}(k_1)P_{T_j|M^J}(t_j|m^J)$ as $K_1$ is independent of $M^J$ and $K_0.$   Reformatting this, we have
$P_{K_1T_j|M^J=m^J}(k_1, t_j)=P_{K_1}(k_1)P_{T_j|M^J=m^J}(t_j)=P_{T_j|M^J=m^J}(t_j)/|\mathcal{K}_1|.$ That is, if we rewrite the joint distribution of $K_1, T_j$ when  $M^J=m^J$, as $P^{m^J}_{K_1T_j}$, then $P^{m^J}_{K_1T_j}=P_{T_j}^{m^J}/|\mathcal{K}_1|.$ Hence, by Theorem \ref{le: code} (property 2),
when $n$ sufficiently large,  we have $ I(K_1,T_j;Z^n|M^J=m^J)\leq{2^{-n\beta_2}}$ for some $\beta_2>0$, where $\beta_2$ does not depend on $m^J.$

Given $M^J=m^J,$ we have that $(K_0, K_1)\rightarrow (T_j, K_1)\rightarrow  Z_j^n$ forms a Markov chain.
Hence, by data processing inequality, we have $I(K_0K_1; Z^n_j|M^J=m^J)\le I(K_1 T_j; Z^n_j|M^J=m^J)\le 2^{-n\beta_2}.$ Thus, averaging over $m^J$, we have $I(K_0 K_1; Z^n_j|M^J)\le I(K_1 T_j; Z_j^n|M^J)\le 2^{-n\beta_2}.$

Further, for any $j\in\{1,\cdots,J\}$ and when  $M^J=m^J$,
\begin{equation*}
Z^n_1\cdots Z^n_{j-1}\rightarrow K_0K_1\rightarrow Z^n_j
\end{equation*}
forms a Markov chain as $X^n_j$ is determined by ($K_0K_1$, $m^J$) and the randomness of sampling $X^   n_j$ from $\mathcal{C}_{K_1T_j}$,  and $Z_j^n$ is determined by $X_j^n$ and the noise in channel $W_2$.
Hence,
\begin{eqnarray}
\nonumber
I(K_0K_1; Z^n_j|Z^n_1\cdots Z^n_{j-1}, M^J=m^J)\\
\nonumber
\leq {I(K_0K_1;Z_j^n|M^J=m^J)}.
\end{eqnarray}
Averaging over $m^J$, we have
\begin{eqnarray}
I(K_0K_1; Z^n_j|Z^n_1\cdots Z^n_{j-1} M^J)
\leq {I(K_0K_1;Z_j^n|M^J)}. \label{eq: chainZJ}
\end{eqnarray}
Hence, by chain rule of mutual information,
\begin{eqnarray*}
&&I(K_0K_1; Z^n_1\cdots Z^n_J M^J)\\
&=&I(K_0K_1; M^J)+I(K_0K_1; Z^n_1\cdots Z^n_J|M^J)\\
&=&I(K_0K_1; Z^n_1\cdots Z^n_J|M^J), \quad(\mbox{$K_0K_1$ is indep of $M^J$})\\
&\leq& \sum_{j}{I(K_0K_1;Z_j^n|M^J)}\le J2^{-n\beta_2}.
\end{eqnarray*}
This concludes our proof.
 \end{IEEEproof}

The second lemma will be used to show that the conditional distribution of secret key on the decision bit is almost uniform.
\begin{lemma} Let $K$ and $V$ be RVs over $\mathcal{K}$ and $\mathcal{V}$ respectively. Then for any $v\in \mathcal{V}$ and any $\mathcal{K}_v\subseteq \mathcal{K}$,
\begin{equation}
{|}P_{K|V=v}{(}\mathcal{K}_v{)}-P_K{(}\mathcal{K}_v{)}{|}\le \frac{1}{2}\textsf{SD}(P_{K|V=v}; P_K).
\end{equation} \label{le: reduc_prob}
\end{lemma}

\begin{IEEEproof}
As $\textsf{SD}(P_{X_1}; P_{X_2})=2\max_{A\subseteq \mathcal{X}} \{P_{X_1}(A)-P_{X_2}(A)\}$ for any RVs $X_1, X_2$ over $\mathcal{X}$,  $P_{K|V=v}(\mathcal{K}_v)-P_{K}(\mathcal{K}_v)\le \frac{1}{2}\textsf{SD}(P_{K|V=v}; P_K).$
Similarly, $-P_{K|V=v}(\mathcal{K}_v)+P_{K}(\mathcal{K}_v)\le \frac{1}{2}\textsf{SD}(P_{K|V=v}; P_K).$
Hence, the lemma follows.
\end{IEEEproof}

The following lemmas will useful in our security proof later.
Lemma \ref{le: SDDel} means that $T'T$ and $K$ are almost independent.
It will be used to assign a value for $\delta_1$ in Theorem \ref{le: code} (3-a).

\begin{lemma} Let   $\{h_{k_0}\}_{k_0\in \mathcal{K}_0}$ be  {\em any} family of functions  from $\mathcal{M}$ to $\mathcal{T}$. Let $K_0, K_1, U, M', M$ be RVs over $\mathcal{K}_0, \mathcal{K}_1, \mathcal{U}, \mathcal{M}$ and $\mathcal{M}$ respectively with $M', M$ being deterministic in $U$. Assume $K_1$ is independent of $(K_0, M)$. Let $T'=h_{K_0}(M')$ and $T=h_{K_0}(M)$. Then, for $\Delta=\textsf{SD}(K_0K_1|U; K_0K_1),$
\begin{eqnarray}
 \textsf{SD}(P_{T'T|K_1}; P_{T'T})\le \sqrt{2\Delta\ln\frac{|\mathcal{K}_0| |\mathcal{K}_1|}{\Delta}}.
\end{eqnarray}
\label{le: SDDel}
\end{lemma}

\begin{IEEEproof}  We have
\begin{eqnarray*}
&&I(T';K_1|T)\le{I(U K_0;K_1|T)}\\
&=& I(U;  K_1|TK_0),\quad  \mbox{ (as $I(K_0;K_1|T)=0$)}\\
&=& H(K_1|TK_0)-H(K_1|U TK_0)\\
&=& H(K_1)-H(K_1|U K_0), \\
&& \mbox{($K_1$ is ind.  of $TK_0$;  $K_0U$ determines $T$)}\\
&=& I(K_1; U K_0)\\
&=& I(K_1; U| K_0),\quad  \mbox{($K_0$ and $K_1$ are ind.)}\\
&\le & I(K_0K_1; U).
\end{eqnarray*}
On the other hand,  by Lemma \ref{le: Ibound},
\begin{eqnarray*}
&& I(T'; K_1|T=t)\\
&\ge& \frac{\left(\sum_{k_1}P_{K_1|T=t}(k_1) \textsf{SD}(P_{T'|T=t}; P_{T'|K_1T=k_1t})\right)^2}{2\ln 2}.
\end{eqnarray*}
By the  convexity of $f(x)=x^2,$  we have
\begin{eqnarray*}
I(T'; K_1|T)=\sum_t P_T(t)I(T'; K_1|T=t)\\
\ge \frac{\left(\sum_{k_1, t}P_{K_1T}(k_1, t) \textsf{SD}(P_{T'|T=t}; P_{T'|K_1T=k_1t})\right)^2}{2\ln 2} .
\end{eqnarray*}
So
\begin{eqnarray*}
&&\sum_{k_1, t}P_{K_1T}(k_1, t) \textsf{SD}(P_{T'|T=t}; P_{T'|K_1T=k_1t})\\
&\le& \sqrt{2I(K_0K_1; U)\ln 2}.
\end{eqnarray*}
Independence between  $K_1$ and $T$  together with Lemma \ref{le: Ibound} gives the  result (after reformatting the left side).
\end{IEEEproof}

The following lemma will be used to assign a value for $\delta_2$ in Theorem \ref{le: code} (3-c) and to show that the  third condition in Theorem \ref{le: code} (3) can be satisfied.

\begin{lemma} Let   $\{h_{k_0}\}_{k_0\in \mathcal{K}_0}$ be  a family of  $\epsilon$-ASU hash functions  from  $\mathcal{M}$ to $\mathcal{T}$. Let $U, M', M$ be RVs over $\mathcal{U}, \mathcal{M}$ and $\mathcal{M}$ respectively s.t. $M', M$ are deterministic in $U$. Let $K_0$ be uniformly random over $\mathcal{K}_0$, $T'=h_{K_0}(M')$ and $T=h_{K_0}(M)$. If $P({M'}=M)=0$, then there exists function $d(t', t)$ s.t. $\sum_{t', t} d(t', t)\le \textsf{SD}(K_0|U; K_0)$ and
\begin{eqnarray*}
P_{T'T}(t', t)\le d(t', t)+\frac{\epsilon}{|\mathcal{T}|}.
\end{eqnarray*}
 \label{le: t't}
\end{lemma}

\begin{IEEEproof}
 Let $\mathcal{K}_0(u, t',  t)$ be the set of $k_0$ so that $h_{k_0}(m')=t'$ and $h_{k_0}(m)=t$, where $m', m$ are the values of $M'$ and $M$ determined by $U=u.$ Let $d(t', t)=\sum_u |P_{K_0U}(\mathcal{K}_0(u, t', t),  u)-P_{K_0}(\mathcal{K}_0(u, t', t))P_{U}(u)|.$  Then,
\begin{eqnarray*}
P_{T'T}(t', t)&=& \sum_uP_{K_0U}(\mathcal{K}_0(u, t', t), u)  \label{eq: close} \\
 &\le & d(t', t)+\sum_u P_{K_0}(\mathcal{K}_0(u, t', t) P_U(u)\\
&\le & d(t', t)+\frac{\epsilon}{|\mathcal{T}|}, \quad \mbox{as $|\mathcal{K}_0(u, t', t)|\le \frac{\epsilon |\mathcal{K}_0|}{|\mathcal{T}|}$}
\end{eqnarray*}

For any $u$,   $\{\mathcal{K}_0({u}, t, t')\}_{t, t'}$ are disjoint. Hence,
\begin{eqnarray}
\nonumber
&&\textsf{SD}(K_0|U; K_0)\\
\nonumber
&=& \sum_{k_0, u}  |P_{K_0U}(k_0, u)-P_{K_0}(k_0)P_U(u)|\\
\nonumber
&\ge &\sum_{t, t',  u}|P_{K_0U}(\mathcal{K}_0(u, t', t),  u)-P_{K_0}(\mathcal{K}_0(u, t', t))P_{U}(u)|\\
\nonumber
&=& \sum_{t, t'} d(t', t).
\end{eqnarray}
This completes the proof.
\end{IEEEproof}

\subsection{Authentication Theorem}

Now we prove our authentication theorem. We need to show that sender Alice can authenticate
 polynomial number of messages using $(K_0, K_1)$, where  the attacker Oscar can  adaptively interleave two types of  attacks.
In Type I attack, when  Alice sends out $(M, X^n)$, Oscar can revise $M$ to $M' (\ne M)$; in Type II attack,   Oscar can send any  pair  $(\hat{M}, \hat{Y}^n)$ to Bob noiselessly. Oscar  succeeds,  if $g_{T'}(Y^n)\in {\cal C}_{K_1T'}$ in a Type I attack (where $T'=h_{K_0}(M')$), or  $g_{\hat{T}}(\hat{Y}^n)\in {\cal C}_{K_1\hat{T}}$ in a Type II attack (where $\hat{T}=h_{K_0}(\hat{M})$).
Our proof idea is as follows.  We use bit $b_\ell=1$ to denote the success of Oscar in the $\ell$th attack (either Type I or Type II).  In a type I attack, there are two cases: (1)  $h_{K_0}(M')=h_{K_0}(M)$ (i.e., $T=T'$), in which case Oscar  succeeds with high probability by the completeness of the coding scheme $(f, g);$ (2) $g_{T'}(Y^n)\in {\cal C}_{K_1T'}$ but $T\ne T'$. For case (1), if $M'$ is independent of $K_0$, then the success of Oscar occurs with probability $\epsilon$ by the property of $h$. Conceivably, if $M'$ is almost independent of $K_0$ (i.e., $\textsf{SD}(K_0|M'; K_0)$ is small), Oscar still succeeds with a small probability. Notice that  $M'$ is determined by the view $U_\ell$ of Oscar. Hence, it suffices to show that  $\textsf{SD}(K_0|U_\ell; K_0)$ is small. For case (2), we can use Theorem \ref{le: code} (3) to show that the success probability is small. For type II attack, if $(\hat{M}, \hat{Y})$  (determined by $U_\ell)$ is independent of $K_1$, then $g_{\hat{T}}(\hat{Y}^n)\in {\cal C}_{K_1\hat{T}}$ holds with probability $\frac{1}{|\mathbb{K}_1|}$.   Conceivably, if $\textsf{SD}(K_1|U_\ell; K_1)$ is small, then this should hold with a small success probability change. As $\textsf{SD}(K_c|U_\ell; K_c)\le \textsf{SD}(K_0K_1|U_\ell; K_0K_1)$ for $c=0, 1,$ we only need to prove that $\textsf{SD}(K_0K_1|U_\ell; K_0K_1)$ is small, which can be  done by properly combining Lemmas \ref{pro:sec}-\ref{le: t't}.

\begin{theorem}
Let $I(X; Y)\ge I(X; Z)+\tau$,  where $Y, Z$ are the outputs of $W_1, W_2$ with input $X$ and   $P_X$ is  a type $P$ with $P(x)>0, \forall x\in \mathcal{X}$. Assume   $h: {\mathcal M}\times{\mathcal K}_0\rightarrow{\mathcal T}$ is  an $\epsilon$-ASU hash function with $\epsilon=\min\{2^{-\Omega(\log n)}, \frac{2^{n^\omega}}{|\mathcal{T}|}\}$ for some $\omega\in (0, 1)$ and $|\mathcal{K}_1|=2^{\Omega(\log n)}$, where $g(n)=\Omega(\log n)$ if $\lim_{n\rightarrow \infty}\frac{g(n)}{\log n}=\infty$.     Then,  our MAC is secure. \label{thm: main}
 \end{theorem}

\begin{IEEEproof}   From Theorem \ref{le: code} (1), the completeness of the MAC holds. Now we concentrate on the authentication property.

Let $M^\nu=M_1\cdots M_\nu$ be the sequence of messages authenticated from  Alice to Bob and $X_i^n, Z_i^n$ be the input and output w.r.t. $M_i$ over  channel $W_2$.  Note  $M^\nu$ is chosen by Alice according to distribution $P_{M^\nu}$ (especially independent of Oscar's random tape $R$);  $X_i^n$ is determined by $(K_0K_1, M_i)$ together with the randomness of sampling $X_i^n$ from $\mathcal{C}_{K_1T}$;  $Z_i^n$ is determined by $X_i^n$ together with the noise in channel $W_2$. It follows that $(M^\nu, K_0K_1, X^n_1Z_1^n\cdots X_\nu^n Z_\nu^n)$ is independent of $R$ and hence has the same distribution as when  Oscar does not present. Hence,
by Lemma \ref{pro:sec},
$I(K_0K_1; M^j Z_1^n\cdots Z_j^n)\le j 2^{-n\beta_2},$ for a constant $\beta_2>0$ and any $j\le \nu$.

As $(M^j, K_0K_1, X^n_1Z_1^n\cdots X_j^nZ_j^n)$ is independent of $R$, $I(K_0K_1; RM^j Z_1^n\cdots Z_j^n)\le j 2^{-n\beta_2}$.  Let  $K\stackrel{def}{=}K_0K_1$ and $V_j\stackrel{def}{=}RM^j Z_1^n\cdots Z_j^n$. By Lemma \ref{lem1},
\begin{eqnarray}
\textsf{SD}(K|V_j; K)\le \sqrt{2j\ln 2}\cdot 2^{-n\beta_2/2}. \label{eq: base}
\end{eqnarray}

According to the adversary model,  Oscar  can adaptively interleave the following attacks.
\begin{itemize}
  \item [I.] When  Alice sends out $(M_j, X_j^n)$, Oscar can revise $M_j$ to $M_j' (\ne M_j)$. He succeeds  if   Bob accepts $(M_j',Y_j^n)$.
\item [II.] At any time,  Oscar can  send a pair  $(\hat{M}, \hat{Y}^n)$ to Bob noiselessly. He succeeds if Bob accepts this pair.
  \end{itemize}
We use bit $b_\ell$ to denote the result of the $\ell$th attack (either type I or type II above) and set $b_\ell=1$ if and only if he succeeds.

 Assume the authentication of  $M^{j_{\ell}-1}$ by Alice has been completed before Oscar launches the $\ell$th attack.   Then, the {\em view}  of Oscar right before the $\ell$th attack  is $U_\ell:=(V_{j_\ell-1}, b_1, \cdots, b_{\ell-1}),$ where  a party's  view is defined as  his random tape $R$ and the data  received externally.

If the $\ell$th attack is Type I, then $b_\ell=1$ iff $g_{T_{j_\ell}'}(Y^n_{j_\ell})\in \mathcal{C}_{K_1T_{j_\ell}'}$ for $T'_{j_\ell}=h_{K_0}(M_{j_\ell}')$.  If we define  event $T'_{j_\ell}=T_{j_\ell}(:=h_{K_0}(M_{j_\ell}))$ by \textsf{col}$_\ell$, and event  $g_{T_{j_\ell}'}(Y^n_{j_\ell})\in \mathcal{C}_{K_1T_{j_\ell}'}$ with  $T_{j_\ell}\ne T_{j_\ell}'$,  by $\textsf{mis}_\ell$, then  $P(b_\ell=1)=P(\textsf{col}_\ell)+P(\textsf{mis}_\ell).$

If the $\ell$th attack is  Type II, then $b_\ell=1$  iff $g_{\hat{T}_\ell}(\hat{Y}^n_\ell)\in \mathcal{C}_{K_1\hat{T}_\ell}$ for $\hat{T}_\ell=h_{K_0}(\hat{M}_\ell),$ where $(\hat{M}_\ell, \hat{Y}^n_\ell)$ is Oscar's output in this attack.

 If $L$ is the upper bound on the number of attacks by Oscar,  then his success probability is
$\Pr\left(\vee_{\ell=1}^L b_{\ell}=1\right).$

As every successful attacker must experience  the first successful attack, we restrict to an attacker who will stop after the first successful attack.
So  $b_\ell=1$ implies $b_1=\cdots=b_{\ell-1}=0.$

Denote the original authentication game by $\Gamma$. Now we modify $\Gamma$ to $\Gamma'$ such that in Type I attack, $b_\ell\stackrel{def}=\textsf{col}_\ell$ (instead of $b_\ell\stackrel{def}{=}\textsf{col}_\ell\vee \textsf{mis}_\ell$).

 Consider an adversary Oscar$'$ for $\Gamma'$  who simply follows the code of   Oscar by setting  each (unknown) \textsf{mis}$_\ell$ as 0 (even if it is 1).  The view of Oscar$'$ in $\Gamma'$ differs from that of Oscar in $\Gamma$ only if $\textsf{mis}_\ell=1$ in $\Gamma'$ for some $\ell$.  Thus,
\begin{equation}
P(succ(\Gamma))\le P(succ(\Gamma'))+\sum_{\ell} P(\textsf{mis}_\ell(\Gamma')). \label{eq: reduc}
 \end{equation}

As $P(succ(\Gamma'))\le \sum_{\ell=1}^L P(b_\ell(\Gamma')=1)$, we only need to bound $P(b_\ell(\Gamma')=1)$ and $P(\textsf{mis}_\ell(\Gamma'))$.  \\

\noindent \underline {\em   Bounding   $P(\textsf{\em mis}_\ell(\Gamma'))$. }   \quad We have the following lemma.

\begin{lemma} $P(\textsf{mis}_\ell(\Gamma'))\le 2^{-\varsigma {n}^\omega}+\sqrt{2\Delta\ln\frac{|\mathcal{K}_0| |\mathcal{K}_1|}{\Delta}}+\Delta$ for a constant $\varsigma>0$, where $\Delta=\textsf{SD}(K|U_\ell; K)$. \label{le: mis}
\end{lemma}

\noindent{\em Proof. }  We first show that  $U_\ell M_{j_\ell}\rightarrow K_1T_{j_\ell}\rightarrow X^n_{j_\ell}\rightarrow Y^n_{j_\ell}$ forms a Markov chain. This follows from two  facts:

(a) \quad Given $X^n_{j_\ell}$, $Y^n_{j_\ell}$ is completely determined by the  noise in channel $W_1$  while this noise  occurs after fixing $(X^n_{j_\ell}, K_1T_{j_\ell}U_{\ell}M_{j_\ell})$ and hence is independent of the latter;

(b) \quad  Given $K_1T_{j_\ell}$, $X_{j_\ell}^n$ is determined by the randomness for sampling   it from $\mathcal{C}_{K_1T_{j_\ell}}$,  which is independent of $U_\ell M_{j_\ell}$.

 By Theorem \ref{le: code}(3) with $\delta_1$ from Lemma \ref{le: t't} and $\delta_2$ from Lemma \ref{le: SDDel}, together with the fact
$\textsf{SD}(K_0|U_\ell; K_0)\le \textsf{SD}(K_0K_1|U_\ell; K_0K_1)$ (from  triangle inequality),  the lemma  follows.  $\hfill\square$ \\

\noindent \underline{\em Bounding $P(b_\ell(\Gamma')=1)$.} \quad Our analysis is for  $\Gamma'$.
Let $\bar{U}_\ell=(V, b_1, \cdots, b_{\ell-1})$, where $V=RM^\nu Z_1^n\cdots Z_\nu^n$. We use $\bar{\mathcal{U}}_\ell^0$ to denote the set of possible values for $\bar{U}_\ell$ with $b_1\cdots b_{\ell-1}=0^{\ell-1}$. So $P(b_\ell=1)=\sum_{u_\ell\in \mathcal{\bar{U}}_\ell^0} P(b_\ell=1, \bar{U}_\ell=u_\ell).$

For given $V=v,$ let    $u_\ell=v|0^{\ell-1}, \ell=1, \cdots, L$.

\vspace{0.10in}\noindent   {\bf Type I attack case: } \quad  In this case,  $b_{\ell}=\textsf{col}_\ell$.  As $M_{j_\ell}', M_{j_\ell}$ are deterministic in $U_{\ell}$ (part of $\bar{U}_\ell$),
\begin{eqnarray*}
\mathcal{E}_{u_\ell}\stackrel{def}{=}\{(k_0, k_1)\in \mathcal{K}: h_{k_0}(M_j')\ne h_{k_0}(M_j)\}
\end{eqnarray*}
 is completely determined  by  $\bar{U}_\ell=u_\ell$.  Hence,
\begin{eqnarray*}
&& \Pr(b_{\ell}=1|\bar{U}_\ell=u_\ell)=P_{K|\bar{U}_\ell=u_\ell}(\mathcal{E}_{u_\ell}^c)\\
\nonumber
&\le& P_{K}(\mathcal{E}_{u_\ell}^c)+\frac{1}{2}\textsf{SD}(P_{K|\bar{U}_\ell=u_\ell}; P_K) \quad (\mbox{by Lemma \ref{le: reduc_prob}}) \\
&\le&  \epsilon+\frac{1}{2}\textsf{SD}(P_{K|\bar{U}_\ell=u_\ell}; P_K).
\end{eqnarray*}
 Averaging over $\bar{U}_\ell$,  $\Pr(b_\ell=1)\le \epsilon+\frac{1}{2}\textsf{SD}(P_{K|\bar{U}_\ell}; P_K).$

\vspace{0.10in}\noindent   {\bf Type II attack case: } \quad  In this case, given  $\bar{U}_\ell=u_\ell$, since  view ${U}_\ell$ of Oscar$'$ is part of $\bar{U}_\ell$, it follows   $(\hat{M}_\ell, \hat{Y}_\ell^n)$ is deterministic in $u_\ell$.  Since   $\mathcal{C}_{\cdot t}$ is a code with decoder $g_t(\cdot)$, $g_{\hat{T}_\ell}(\hat{Y}^n_\ell)\in \mathcal{C}_{K_1\hat{T}_\ell}$ holds for at most one $K_1$ when $K_0$ and $u_\ell$ are fixed. Thus, given $\bar{U}_\ell=u_\ell$,  $b_\ell=1$ holds for at most $|\mathcal{K}_0|$ choices of $(K_0, K_1).$  Let $\mathcal{E}_{u_\ell}=\{(k_0, k_1): g_{\hat{T}_\ell}(\hat{Y}_\ell^n)=\perp\}$. Then
\begin{eqnarray*}
&& \Pr(b_{\ell}=1|\bar{U}_\ell=u_\ell)\\
&\le& P_{K|\bar{U}_\ell=u_\ell}(\mathcal{E}_{u_\ell}^c)\\
&\le&  \frac{1}{|\mathcal{K}_1|}+\frac{1}{2}\textsf{SD}(P_{K|\bar{U}_\ell=u_\ell}; P_K) \quad (\mbox{by Lemma \ref{le: reduc_prob}})
\end{eqnarray*}
 Averaging over $\bar{U}_\ell$, $ \Pr(b_\ell=1)\le \frac{1}{|\mathcal{K}_1|}+\frac{1}{2}\textsf{SD}(P_{K|\bar{U}_\ell}; P_{K}).$

\vspace{.10in} \noindent\underline{\em Bounding $\textsf{\em SD}(P_{K|\bar{U}_\ell}; P_{K})$. } \quad Given  $\bar{U}_\ell=u_\ell=v0^{\ell-1}$, we  must have $K\in \mathcal{E}_{u_i}$ for any $i<\ell$.   So $K\in \mathcal{K}_{v}^\ell\stackrel{def}=\cap_{i=1}^{\ell-1} \mathcal{E}_{u_i}.$  In Type I attack, $b_\ell$ in $\Gamma'$ is determined by $(K_0, M_{j_\ell}', M_{j_\ell})$, which is further determined by $(K_0, V_{j_\ell}, b_1, \cdots, b_{\ell-1}).$
In Type II attack, $b_\ell$ in $\Gamma'$ is determined by $(K, \hat{M}_\ell, \hat{Y}_\ell)$, which is further determined by $(K, V_{j_\ell}, b_1, \cdots, b_{\ell-1}).$ It follows that $(b_1, \cdots, b_{\ell})$ is deterministic in $(K, V).$ As $\bar{U}_\ell=(V, b_1, \cdots, b_{\ell-1})$,  from rule $P_{AB}=P_AP_{B|A}$, we have
$P_{K\bar{U}_\ell}(k, u_\ell)=P_{KV}(k, v)$ if $(b_1,\cdots, b_{\ell-1})$ determined by $(k, v)$ is $0^{\ell-1}$; 0 otherwise.  Note $\mathcal{K}_v^\ell$ is the set of all possible $k$ such that $(b_1,\cdots, b_{\ell-1})$ determined by $(k, v)$ is $0^{\ell-1}$. Thus,
\begin{eqnarray}
P_{\bar{U}_\ell}(u_\ell)=\sum_{k\in \mathcal{K}_v^\ell} P_{KV}(k, v)=P_{KV}(\mathcal{K}_v^\ell, v).
\end{eqnarray}
  Hence,
\begin{align*}
 &\textsf{SD}(P_{K|\bar{U}_\ell}; P_{K})\\
=&\sum_v \sum_{k\in \mathcal{K}_v^{\ell}}|P_{KV}(k, v)-P_{KV}(\mathcal{K}_v^\ell, v)P_{K}(k)|\\
&+\sum_v \sum_{k\not \in \mathcal{K}_v^{\ell}}|P_{KV}(\mathcal{K}_v^\ell, v)P_{K}(k)|\\
 \le& \textsf{SD}(K|V; K)+2\sum_v P_{KV}(\mathcal{K}\backslash\mathcal{K}_v^\ell, v)\\
\le&  2 \textsf{SD}(K|V; K)+2\sum_v P_{K}(\mathcal{K}\backslash\mathcal{K}_v^\ell) P_V(v), (\mbox{Lemma \ref{le: reduc_prob}}) \\
\le & 2\textsf{SD}(K|V; K)+2(\ell-1)\epsilon',
\end{align*}
where $\epsilon'=\max(\epsilon, \frac{1}{|\mathcal{K}_1|}).$

\vspace{.10in} \noindent \underline{\em Finalizing the bound on  $P(succ(\Gamma))$. }\quad
As $U_\ell$ is part of $\bar{U}_\ell$, it follows that
$\textsf{SD}(K|U_\ell; K)\le \textsf{SD}(K|\bar{U}_\ell; K).$ Notice $\textsf{SD}(K|V;K)\le \sqrt{2\nu\ln2} \cdot 2^{-n\beta_2/2}$. By Lemma \ref{le: mis} and calculus analysis, there exists  $\varsigma'>0$ and $\omega'<\omega$  such that $P(\textsf{mis}_\ell(\Gamma'))$ is bounded by
\begin{eqnarray*}
 2^{-\varsigma' {n}^{\omega'}}+\sqrt{4(\ell-1)\epsilon'\ln\frac{|\mathcal{K}_0| |\mathcal{K}_1|}{2(\ell-1)\epsilon'}}+2(\ell-1)\epsilon'.
\end{eqnarray*}
Summarizing the  bound on $P(b_\ell=1)$, we have $P(b_\ell=1)\le \sqrt{2\nu\ln2} \cdot 2^{-n\beta_2/2}+\ell \epsilon'.$

As $P(succ(\Gamma'))\le \sum_\ell P(b_\ell(\Gamma')=1)$ and $\nu$ is polynomially bounded, Eq. (\ref{eq: reduc}) gives
 \begin{eqnarray*}
&&P(succ(\Gamma))\\
&\le& \sum_\ell P(\textsf{mis}_\ell(\Gamma'))+\sum_\ell P(b_\ell(\Gamma')=1)\\
&\le& 2^{-\varsigma''{n}^{\omega'}}+\sum_{\ell=1}^L (\sqrt{4(\ell-1)\epsilon'\ln\frac{|\mathcal{K}_0| |\mathcal{K}_1|}{2(\ell-1)\epsilon'}}+3\ell\epsilon')\\
&\le& 2^{-\varsigma''{n}^{\omega'}}+2L\sqrt{L\epsilon'\ln\frac{|\mathcal{K}_0| |\mathcal{K}_1|}{\epsilon'}}+{3L^2\epsilon'},
 \end{eqnarray*}
for some $\varsigma''>0$.  This is negligible as $L$ is polynomial in $n$ and $\epsilon'$ is negligible. This completes our theorem.
\end{IEEEproof}

\section{Efficiency}
The following definitions are defined in the introduction and we repeat them here  for convenience. In the proposed MAC scheme, the authentication rate $\rho_{auth}$ can be rewritten as $\rho_{auth}=\rho_{tag}\cdot\rho_{chan}$, where $\rho_{tag}=\frac{\log |\mathcal{M}|}{\log |\mathcal{T}|}$  and  $\rho_{chan}=\frac{\log |\mathcal{T}|}{n}$. We call  $\rho_{tag}$ the  {\em tag rate} and  $\rho_{chan}$ the  {\em channel coding rate}.  Tag rate $\rho_{tag}$ is mainly determined by cryptographic techniques.

For our construction, the constraint for $\mathcal{T}$ is $\mathcal{T}\subset [{\mathbb J}]$. The constraint for ${\mathbb J}$ is $\frac{\log {\mathbb J}}{n}<H(X|Y)+\tau$ (Theorem \ref{le: code}), where $\tau$ only has the constraint $H(X|Z)> H(X|Y)+\tau$ (Theorem \ref{le: code} and Theorem 2). So for any  $\delta\in (0, H(X|Z)-H(X|Y))$, we can define $\tau=H(X|Z)-H(X|Y)-\delta/2$ and then set   $|\mathcal{T}|={\mathbb J}=2^{n(H(X|Z)-\delta)}$.       Under this, $\frac{\log |\mathcal{T}|}{n}=H(X|Z)-\delta$.  We can summarize this observation as follows.  \\

\begin{corollary} Keep  conditions in Theorem \ref{thm: main}. For any $\delta\in (0, H(X|Z)-H(X|Y))$, let  $\tau=H(X|Z)-H(X|Y)-\delta/2$. Then our MAC is secure with  $\rho_{auth}=[H(X|Z)-\delta]\cdot\rho_{tag}$.  \label{co: main}
\end{corollary}

\subsection{Comparison with A Natural  Scheme} In our MAC construction, we first compute $T$ and then encode it to $X^n$ using the code in Theorem \ref{le: code}. A natural variant scheme   is similar to ours, except that $T$ is encoded to $X^n$ using the classic secrecy code of Csisz\'{a}r and K\"{o}rner \cite{Csiszar1}, where the decoding is simply to decode $T$ and check its consistency with $M'$. The security of this scheme is straightforward as $T$ is fully protected. Let the secrecy capacity of the wiretap  channel $(W_1, W_2)$ is $C_s$. Then the authentication rate of this scheme is $\rho_{auth}=\rho_{tag} C_s.$ According to  \cite{Csiszar1}, if  channel $W_1$ is less noisy than channel $W_2$, then $C_s=H(X|Z)-H(X|Y)$. Under this, the ratio of the authentication rate of this scheme  to ours is  arbitrarily close to  ${1-H(X|Y)/H(X|Z)}<1$ (as $\delta$ can be arbitrarily small).

The above observation is surprising. Indeed, since $T$ in the natural scheme is encoded using the capacity achieving code, the above comparison  seems to signify that our MAC does not protect $T$ in its  full secrecy because we have achieved  a higher rate. Our  explanation for this  is as follows. The secrecy capacity of a wiretap channel has two tasks: (a) the adversary has no information about the secret message; (b) the legal receiver Bob  should be able to recover the secret message. In our setting, we  only need to handle task (a) but not (b), as  Bob can recover $T$ from $M'$ in the noiseless channel (if $M'=M$) while when $M'\ne M$, his  job is only to realize and reject the authentication. So in our scheme, there is no guarantee that Bob can recover $T$ from $Y^n.$

\subsection{Realization of our MAC}

To realize  our scheme, we only need to specify $h_k$ and $\mathcal{K}_0, \mathcal{K}_1$ and $\tau.$ Let $\tau=H(X|Z)-H(X|Y)-\delta/2$ as in Corollary \ref{co: main}. Then, $\rho_{auth}=[H(X|Z)-\delta]\cdot\rho_{tag}.$
Further, we  realize  $h_k$ with  $\frac{s}{q}$-ASU in Lemma \ref{le: stin}, where $|\mathcal{K}_0|=q^{s}$, $|\mathcal{M}| = q^{2^s}$
and $|\mathcal{T}|= q$.   Let $|\mathcal{K}_1|=2^{\log^2 n}.$
 It is easy to verify that under this setup,  the security condition in  our authentication theorem is satisfied as long as $s<2^{n^{\omega}}$ for some $\omega\in (0, 1)$. As a result,   $\rho_{tag}=2^{s}$ and hence $\rho_{auth}=[H(X|Z)-\delta]2^s$, where $s<2^{n^{\omega}}$ for some $\omega\in (0, 1)$.

\section{Conclusion}

We considered an  authentication problem, in which Alice authenticated a source $M$ over a wiretap channel $(W_1, W_2)$ under the assistant of a noiseless channel. Alice and Bob shared a secret key.  We studied the framework, where Alice sent the insecure information  $S$ over the noiseless channel and an encoded tag $T$ over the wiretap channel. We proposed an efficient MAC scheme  for wiretap channel $(W_1,W_2)$, in which the authentication rate beat the intuitively best possible result. An immediate open problem is how to construct a computationally efficient protocol (instead of channel efficient one  studied in this paper).

 \section*{Acknowledgments}
 \noindent Shaoquan Jiang would like to thank Huaxiong Wang for suggesting  this topic. This work  is supported by National 973 Program of China (No. 2013CB834203), NSFC (No. 60973161, No. 61133016) and the National High Technology Joint Research Program of China (863 Program, Grant No. 2011AA010706).

\appendix

In this Appendix we provide the proof of Theorem 1.

\subsection{Preparation}

Let $X, Y$ be RVs over $\mathcal{X}$ and $\mathcal{Y}$ respectively with  a joint distribution $P_{XY}.$ Let $(X^n,Y^n)$ be $n$ independent  outputs according to   $P_{XY}$. In this case,  $(X^n,Y^n)$ is called   a {\em discrete memoryless multiple source} ({DMMS}) with generic variables $X, Y$. For $\mathcal{A}\subseteq {\mathcal{X}}^n$, let  $\widetilde{P}_{X^nY^n}$ be the
joint distribution of $(X^n, Y^n)$, conditional on $X^n\in \mathcal{A}$. That is,  $\widetilde{P}_{X^nY^n}(x^n,y^n)\stackrel{def}{=}P^n_{XY}(x^n,y^n)/P_X^n(\mathcal{A})$ for any $x^n\in{\mathcal{A}}, y^n\in{{\mathcal{Y}}^n}.$
Marginal distributions  $\widetilde{P}_{X^n}(x^n)=\sum_{y^n\in \mathcal{Y}^n}P^n_{XY}(x^n,y^n)/P_X^n(\mathcal{A})=P_{X}^n(x^n)/P_X^n(\mathcal{A})$ and  $\widetilde{P}_{Y^n}(y^n)=\sum_{x^n\in \mathcal{A}}P^n_{XY}(x^n,y^n)/P_X^n(\mathcal{A})$.

For any index set  $\mathcal{B}$, any collection of disjoint subsets $\{\mathcal{A}_b\}_{b\in \mathcal{B}}$ with $\cup_{b\in \mathcal{B}}\mathcal{A}_b=\mathcal{A}$ forms a partition of $\mathcal{A}.$  Of course, a partition of $\mathcal{A}$ does not depend on the index set $\mathcal{B}$. The  generality of $\mathcal{B}$ is only for our ease of presentation.

For  a partition  $\{\mathcal{A}_b\}_{b\in \mathcal{B}}$ of $\mathcal{A}$, let $\widetilde{P}_{Y^n|b}(y^n)\stackrel{def}{=} \widetilde{P}(Y^n=y^n|X^n\in \mathcal{A}_b).$ That is,
\begin{eqnarray}
\widetilde{P}_{Y^n|b} (y^n)&=&\sum_{x^n\in \mathcal{A}_b}{{\widetilde{P}_{X^nY^n}(x^n,y^n)}/{\widetilde{P}_{X^n}({\mathcal{A}}_b)}}\\
&=&\sum_{x^n\in \mathcal{A}_b} P^n_{XY}(x^n, y^n)/P_X^n(\mathcal{A}_b).
\end{eqnarray}
In other words, $\widetilde{P}_{Y^n|b}$ equals the marginal distribution of $Y^n$ in $P^n_{XY}$, conditional on $X^n\in \mathcal{A}_b.$

A partition can also be characterized through  a mapping. Specifically, for mapping  $\sigma:\mathcal{A}\rightarrow \mathcal{B}$, let $\mathcal{A}_b\stackrel{def}{=}\sigma^{-1}(b)$ for $b\in \mathcal{B}$. Then $\{\mathcal{A}_b\}_{b\in \mathcal{B}}$  forms  a partition of $\mathcal{A}$. On the other hand, given a partition $\{\mathcal{A}\}_{b\in \mathcal{B}}$, we can define $\sigma: \mathcal{A}\rightarrow \mathcal{B}$ by $\sigma(x)=b$ for all $x\in \mathcal{A}_b.$ Thus, when the context is clear,  we will simply call  a mapping \emph{$\sigma$ a partition of size $|\mathcal{B}|$ for $\mathcal{A}$}.

For any partition $\sigma:\mathcal{A}\rightarrow \mathcal{B}$, $\sigma(X^n)$ has a distribution induced by random variable $X^n$. As $\sigma(x^n)=b$ if and only if $x^n\in \mathcal{A}_b$, we have  $\Pr(\sigma(X^n)=b)=\widetilde{P}_{X^n}(\mathcal{A}_b)=P_X^n(\mathcal{A}_b)/P_X^n(\mathcal{A}).$
Thus, under $\widetilde{P}_{X^nY^n}$ for $(X^n, Y^n)$,
\begin{eqnarray}
\nonumber
&&\textsf{SD}(Y^n|\sigma(X^n);Y^n)\\
\nonumber
&=&
\sum_{b\in \mathcal{B}}\widetilde{P}_{X^n}(\mathcal{A}_b)\sum_{y^n\in{\mathcal{Y}^n}}{|\widetilde{P}_{Y^n|b}(y^n)-\widetilde{P}_{Y^n}(y^n)|}\\
&=&\sum_{b\in \mathcal{B}}\widetilde{P}_{X^n}(\mathcal{A}_b) \textsf{SD}(\widetilde{P}_{Y^n|b};\widetilde{P}_{Y^n}). \label{eq: sd}
\end{eqnarray}

If $P_X=P$ for a type $P$ and $\mathcal{A}=\textsf{T}_P^n$,  Csisz\'{a}r \cite{Csiszar2} showed that when $k$ is not too large, there exists a partition $\sigma$ that partitions  $\textsf{T}_{P}^n$ into $k$ subsets of  almost equal size  so that $\sigma(X^n)$  is almost independent of  $Y^n$. This is the following.

\begin{lemma}\label{pro: partition}
\cite{Csiszar2}  DMC $W: \mathcal{X}\rightarrow \mathcal{Y}$ has input $X$ and output $Y$, where $X$ is according to a type $P$ with  $P(x)>0, \forall x\in \mathcal{X}$. Then, for any $\tau>0$, there exists $\beta>0$ such that when   $n$ large enough and $k\le |\textsf{T}_P^n|2^{-n(I(X; Y)+\tau)}$, $\textsf{T}_P^n$ has a partition $\sigma: \textsf{T}_P^n\rightarrow \{1, \cdots, k\}$ satisfying
\begin{equation}\label{eq6}
|\mathcal{A}_i|=\frac{|\textsf{T}_P^n|}{k}(1+\epsilon_i), ~~~~~~~~ \textsf{SD}(Y^n|\sigma(X^n); Y^n)<2^{-n\beta},
\end{equation}
where $\mathcal{A}_i=\sigma^{-1}(i)$ and $|\epsilon_i|\le 2^{-n\beta}.$
Moreover, if $\sigma$ is  uniformly random  among  all possible partitions, then Eq.  (\ref{eq6}) holds, except for an exponentially small (in $n$)  probability.
\end{lemma}

{\em Remark. } This lemma can be trivially generalized to the setting $\sigma': \textsf{T}_P^n\rightarrow \mathcal{B}$ with $|\mathcal{B}|=k$ as the result does not depend on the choice of $\mathcal{B}$. Specifically, for any $\sigma$ in the lemma and $\mathcal{B}=\{b_1, \cdots, b_k\}$, define $\sigma'=\sigma\circ \pi$, where mapping $\pi: \mathcal{B}\rightarrow \{1, \cdots, k\}$ with $\pi(b_i)\mapsto i$ is one-one.  With  $\mathcal{A}_{b_i}=\mathcal{A}_i$, $\sigma$ satisfies Eq. (\ref{eq6}) if and only if $\sigma\circ \pi$ satisfies the corresponding properties with index set $\mathcal{B}$. Later, we will  reference this lemma for a general $\mathcal{B}$ without a justification.

\vspace{.05in}  For any set $\mathcal{A}$, there are $k^{|\mathcal{A}|}$ partitions of size $k$.  One can sample a uniformly random partition $\sigma: \mathcal{A}\rightarrow \mathcal{B}$ by assigning $\sigma(x)$ to a uniformly random element $b$ in $\mathcal{B}$ for each $x\in \mathcal{A}, b\in \mathcal{B}.$ This view will be used in the following theorem.

\subsection{Useful lemmas}

Now, we present  some  lemmas that will be used to prove   theorem 1 later.  The first lemma bounds $E[|(\textsf{T}_{[W]_\epsilon}(Z_1^n)\cap \textsf{T}_{[W]_\epsilon}(Z_1^n)|]$ for randomly chosen $Z_1^n, Z_2^n$ with type $P_X.$  Our idea is to notice that for a random subset $B$ of $S$,
$E(|B|)=\sum_{y\in S} P(y\in B).$ So we only need to bound
\begin{align}
&\sum_{y^n} P(y^n\in \textsf{T}_{[W]_\epsilon}(Z_1^n)\cap \textsf{T}_{[W]_\epsilon}(Z_1^n)). \label{eq: intersec-P1}
\end{align}
It is easy to show that $y^n\in \textsf{T}_{[W]_\epsilon}(Z^n)$ for a typical $Z^n$ ($Z^n$ with type $P_X$ satisfies this)    implies  $Z^n\in \textsf{T}^n_{[X|Y]_\epsilon}(y^n).$ So Eq. (\ref{eq: intersec-P1}) is bounded by $\sum_{y^n} P(Z_1^n, Z_2^n\in \textsf{T}^n_{[X|Y]_\epsilon}(y^n)).$  Notice that $Z_1^n, Z_2^n$ are independent and $P(Z^n\in \textsf{T}^n_{[X|Y]_\epsilon}(y^n))\approx 2^{-nI(X; Y)}.$ The desired  bound for Eq. (\ref{eq: intersec-P1}) can be obtained by direct calculations.

\begin{lemma}  Assume RVs $X$ and $Y$ are connected by DMC $W: \mathcal{X}\rightarrow \mathcal{Y}$ where $P_X=P$ for some type $P$.  Let $(Z_1^n, Z_2^n)$ be a uniformly randomly pair from  $\textsf{T}^n_P$.   Then, there exists a constant $c>0$ such that for  any $\epsilon>0$,  when $n$ large enough,
\begin{eqnarray}
E\Big{(}|\textsf{T}_{[W]_\epsilon}^n({Z}^n_1)\cap \textsf{T}_{[W]_\epsilon}^n({Z}^n_2)|\Big{)}\le 2^{n[H(Y|X)-I(X; Y)+c\epsilon]}.
\end{eqnarray}  \label{le: intersec}
\end{lemma}

\begin{IEEEproof}
  For a fixed  set $S$ and its  random subset  $B\subseteq S$, $E(|B|)=E(\sum_{y\in S} {\bf 1}_B(y))=\sum_{y\in S}P(y\in B),$ where ${\bf 1}_B(y)=1$ if $y\in B$ and 0 otherwise. Thus,
\begin{eqnarray}
&&E\Big{(}|\textsf{T}_{[W]_\epsilon}^n({Z}^n_1)\cap \textsf{T}_{[W]_\epsilon}^n({Z}^n_2)|\Big{)}\\
&=& \sum_{y^n\in \mathcal{Y}^n} P(y^n\in \textsf{T}_{[W]_\epsilon}^n({Z}^n_1)\cap \textsf{T}_{[W]_\epsilon}^n({Z}^n_{2})). \label{eq: cont-1}
\end{eqnarray}
Notice that  $y^n\in \textsf{T}_{[W]_\epsilon}(x^n)$ for $x^n\in \textsf{T}_{P}^n$ implies
\begin{eqnarray*}
|P_{x^ny^n}(a, b)-P_X(a){P}_{Y|X}(b|a)|\le \frac{\epsilon}{|\mathcal{X}||\mathcal{Y}|}
\end{eqnarray*}
 for all $a, b$ as  $P_X=P$. Summation over $a$ implies
 \begin{eqnarray}
|P_{y^n}(b)-P_Y(b)|\le \frac{\epsilon}{|\mathcal{Y}|}. \label{eq: TY}
\end{eqnarray}
 This further implies that  $|P_{x^ny^n}(a, b)-P_{y^n}(b)P_{X|Y}(a|b)|\le \frac{c'\epsilon}{|\mathcal{X}||\mathcal{Y}|}$ for some constant $c'>0$. So for   $x^n\in \textsf{T}_P^n$, $y^n\in \textsf{T}_{[W]_\epsilon}(x^n)$ implies $x^n\in \textsf{T}^n_{[X|Y]_{c'\epsilon}}(y^n).$  It follows that $\{x^n\in \textsf{T}_P^n: y^n\in \textsf{T}_{[W]_\epsilon}^n(x^n)\}\subseteq \{x^n\in \textsf{T}_P^n: x^n\in \textsf{T}^n_{[X|Y]_{c'\epsilon}}(y^n)\}\subseteq \textsf{T}_{[X|Y]_{c'\epsilon}}^n(y^n),$ which has a size at most $2^{n[H(X|Y)+c''\epsilon]}$ for some constant $c''>0$  by  Lemma \ref{le: basicH} (3). So Eq. (\ref{eq: cont-1}) gives
\begin{eqnarray*}
&&\sum_{y^n\in \mathcal{Y}^n} P(y^n\in \textsf{T}_{[W]_\epsilon}^n({Z}^n_1)\cap \textsf{T}_{[W]_\epsilon}^n({Z}^n_2))\\
 &=& \sum_{y^n\in \textsf{T}_{[Y]_\epsilon}^n} P(y^n\in \textsf{T}_{[W]_\epsilon}^n({Z}^n_1)\cap \textsf{T}_{[W]_\epsilon}^n({Z}^n_2)) (\mbox{by Eq. (\ref{eq: TY})})\\
&\le&  \sum_{y^n\in \textsf{T}_{[Y]_\epsilon}^n} P(Z_1^n, Z_2^n \in \textsf{T}_{[X|Y]_{c'\epsilon}^n}(y^n)) \\
&\stackrel{*}{\le} &\sum_{y^n\in \textsf{T}_{[Y]_\epsilon}^n} \frac{2^{n[H(X|Y)+c''\epsilon]}}{|\textsf{T}_P^n|}\times \frac{2^{n[H(X|Y)+c''\epsilon]}}{|\textsf{T}_P^n|-1}\\
&\le&2(n+1)^{2|\mathcal{X}|}\sum_{y^n\in \textsf{T}_{[Y]_\epsilon}^n} 2^{-2n[I(X; Y)-c''\epsilon]} \ (\mbox{Lemma \ref{le: basicH}(1)}) \\
&\le&2^{-n[I(X; Y)-H(Y|X)-(2c''+c^*+1)\epsilon]} \ (\mbox{Lemma \ref{le: basicH}(2)}),
\end{eqnarray*}
for some $c^*>0.$ Ineq (*) holds as $Z_1^n, Z_2^n$ is a uniformly random pair in $\textsf{T}_P^n$. The lemma holds  with $c=2c''+c^*+1$.  \end{IEEEproof}

The second lemma essentially states that if we randomly  sample a subset $A$ of size at most $2^{n(I(X;Y)-\tau)}$ from $\textsf{T}_P^n$ for some $\tau>0$, then   most likely $A$ is an error-correcting code with  an exponentially small error.  The basic idea is simple. By the previous lemma, if the sampled set is $\{Z_1^n, \cdots, Z_\ell^n\}$, then $\textsf{T}_{[W]_\epsilon}^n(Z_i^n)\cap\cup_{j\ne i} \textsf{T}_{[W]_\epsilon}^n(Z_j^n)$ has a  size of  {\em roughly} $2^{n(H(Y|X)-\tau)}$, which is an exponentially small part  of  $\textsf{T}_{[W]_\epsilon}^n(Z_i^n).$ So $A$ is a code under a typical decoding that has an  exponentially small error. The formal proof is to make the above rough idea rigorous through probability arguments.

\begin{lemma} Let $P$ be a type over $\mathcal{X}$. Assume integer   $\ell\le  2^{n(I(X; Y)-\tau)}$ for some $\tau>0$ and RVs $X$ and $Y$ are  connected by DMC $W: \mathcal{X}\rightarrow \mathcal{Y}$ with $P_X=P$.  Let $A:=\{Z_1^n, \cdots, Z_\ell^n\}$ (indexed randomly) be a purely random subset of $\textsf{T}_P^n$ of size $\ell$.  Let $(f, g)$ be a code with codebook $A$, where encoding $f: [\ell]\rightarrow A, f(i)\mapsto Z^n_i$, and  decoding $g(Y^n)=i$ if there exists a  unique $i$ s.t. $Y^n\in \textsf{T}_{[W]_\epsilon}^n(Z_i^n)$ and $g(Y^n)=\perp $ otherwise.   Then, there exist constants $\lambda>0, \epsilon_0>0$ (not depending on $\ell$) such that   with probability at least $1-2^{-n\tau/2}$ (over the choice of  $A$), we have   $e(A)\le 2^{-n\lambda\epsilon^2}$,  for  any  $\epsilon<\epsilon_0$ and when $n$ large enough.     \label{le: pack}
\end{lemma}

\begin{IEEEproof}
We first compute
\begin{eqnarray*}
\nonumber
&& E(|\textsf{T}_{[W]_\epsilon}^n({Z}^n_i)\cap \cup_{j\ne i} \textsf{T}_{[W]_\epsilon}^n({Z}^n_j)|)\\
\nonumber
&\le & \sum_{j\in [\ell]\backslash \{i\}} E\Big{(}|\textsf{T}_{[W]_\epsilon}^n({Z}^n_i)\cap \textsf{T}_{[W]_\epsilon}^n({Z}^n_j)|\Big{)}\\
\nonumber
&\le & \ell\cdot  E\Big{(}|\textsf{T}_{[W]_\epsilon}^n({Z}^n_1)\cap \textsf{T}_{[W]_\epsilon}^n({Z}^n_2)|\Big{)}\\
\nonumber
&&\mbox{($Z_1^n, \cdots, Z_\ell^n$ are symmetric}) \\
&\le&  \ell\cdot 2^{-n[I(X; Y)-H(Y|X)-c\epsilon]} \quad \mbox{(by Lemma \ref{le: intersec})}\\
&\le & 2^{n[H(Y|X)-\tau+c\epsilon]},\quad  \mbox{$n$ large enough}
\end{eqnarray*}
for some constant $c>0.$
Hence,
\begin{eqnarray*}
\sum_{i=1}^n \frac{1}{\ell} E(|\textsf{T}_{[W]_\epsilon}^n({Z}^n_i)\cap \cup_{j\ne i} \textsf{T}_{[W]_\epsilon}^n({Z}^n_j)|)\le 2^{n[H(Y|X)-\tau+c\epsilon]}.
\end{eqnarray*}
By Markov inequality,  with probability $1-2^{-n\tau/2}$ (over $A$),
\begin{eqnarray*}
\sum_{i=1}^n \frac{1}{\ell} |\textsf{T}_{[W]_\epsilon}^n({Z}^n_i)\cap \cup_{j\ne i} \textsf{T}_{[W]_\epsilon}^n({Z}^n_j)|\le 2^{n[H(Y|X)-\frac{\tau}{2}+c\epsilon]}.
\end{eqnarray*}
Denote the collection of such $A$ by ${\cal A}$.

By Lemma \ref{le: basicH} (3), there exists constant $\hat{c}>0$ s.t.   $P_{Y|X}^n(y^n|x^n)\le 2^{-n[H(Y|X)-\hat{c}\epsilon]},$  $\forall \epsilon>0, \forall x^n\in \textsf{T}_P^n, \forall y^n\in \textsf{T}_{[W]_\epsilon}(x^n)$. So there exists constant $c'>0$ s.t. for any $A\in {\cal A}$, when $I$ is uniformly random in $[\ell]$,
 \begin{eqnarray}
P\Big{(}Y^n\in \textsf{T}_{[W]_\epsilon}^n(Z^n_I)\cap \cup_{j\ne I} \textsf{T}_{[W]_\epsilon}^n({Z}^n_j)\Big{)}\le 2^{-n(\tau/2-c'\epsilon)}.
\end{eqnarray}
 Note an error occurs only if $Y^n\in \textsf{T}_{[W]_\epsilon}^n(Z^n_I)\cap \cup_{j\ne I} \textsf{T}_{[W]_\epsilon}^n({Z}^n_j)$ or if $Y^n\not\in \textsf{T}_{[W]_\epsilon}^n({Z}^n_I).$ Thus, by Lemma \ref{le: basicH} (4), there exists constant $\lambda_0>0$ s.t.
   $e(A)\le 2^{-n(\tau/2-c'\epsilon)}+2^{-n\lambda_0\epsilon^2}$.  Lemma follows with  $\lambda<\lambda_0$ and $\epsilon_0$ small enough (dependent on $\tau, c', \lambda_0$).
\end{IEEEproof}

The lemma below states that a random subset ${A}$ of  $\textsf{T}_P^n$ with $|{A}|=\ell$ is  uniformly random over all subsets with  size $\ell$.
\begin{lemma} For a type $P$ and integer $s$,   a subset ${A}\subseteq \textsf{T}_P^n$ is sampled by including each $x^n\in \textsf{T}_P^n$ with probability $1/s$. Then
given $|A|=\ell$, $A$ is uniformly random over all possible subsets of $\textsf{T}_P^n$ of size $\ell.$ \label{le: uniform}
\end{lemma}

\emph{Proof.} Let $N=|\textsf{T}_P^n|.$ Then, a particular set $A$ of size $\ell$ is sampled with probability $s^{-\ell}(1-1/s)^{N-\ell}$, which does not depend on the specific element of $A$.    So given $|A|=\ell$, $A$ occurs with probability ${1}/{{N\choose \ell}}.$ $\hfill\square$ \\

For $s<|\textsf{T}_P^n|2^{-n(I(X;Y)-\theta)}$, in the following lemma, we want to claim that for a random partition $\mathcal{A}_1, \cdots, \mathcal{A}_s$ of  $\textsf{T}_P^n$, with high probability, most of $\mathcal{A}_j$'s  are  codes.
Our proof strategy is mainly to repeatedly use  the following fact: if $E(X)\le L$ for $L>0$ and RV $X$, then $P(X>uL)\le 1/u$ for any $u>0.$ This fact is a simple consequence of Markov inequality.

\begin{lemma} Let RVs $X, Y$ be connected by DMC $W$. For a type $P$ and $s=|\textsf{T}_P^n|2^{-n(I(X;Y)-\theta)}$,  $\mathcal{A}_1, \cdots, \mathcal{A}_s$ is  a random partition  of $\textsf{T}_P^n$:  for each $x^n\in \textsf{T}_P^n$, take a uniformly random $i\in [s]$ and put $x^n$ into $\mathcal{A}_i$. Regard  $\mathcal{A}_j$ with  $|\mathcal{A}_{j}|\le 2^{I(X;Y)-\theta/2}$ as  a  code in Lemma \ref{le: pack} and $\mathcal{A}_j$ with  $|\mathcal{A}_{j}|>2^{I(X; Y)-\theta/2}$ as  a code of error 1.  Then, there exist  constants $\lambda>0, \epsilon_0>0$  such that,
with probability $1-2^{-n\theta/8+1}$ (over the randomness of partition), there are at most $2^{-n\theta/8}s$ possible $j$'s with  $e(\mathcal{A}_{j})> 2^{-n\lambda\epsilon^2}$,  for any $\epsilon<\epsilon_0$. \label{le: bound}
\end{lemma}

\begin{IEEEproof}
 By Lemma \ref{le: uniform}, given $|\mathcal{A}_{j}|=\ell$, $\mathcal{A}_{j}$ is uniformly random over all possible subsets of $\textsf{T}_P^n$ of  size $\ell$.  So by Lemma \ref{le: pack}, given  $|\mathcal{A}_{j}|=\ell\le 2^{I(X;Y)-\theta/2}$, there exist  constants $\lambda>0$ and  $\epsilon_0>0$ (not depending on $\ell$) such that,  with probability $1-2^{-n\theta/4}$,  $\mathcal{A}_{j}$ is  a code   with
\begin{eqnarray}
e(\mathcal{A}_{j})\le 2^{-n\lambda\epsilon^2}, \label{eq: Ajgoo}
\end{eqnarray}
 for  any $\epsilon<\epsilon_0$. Here by symmetry of $\mathcal{A}_{1}, \cdots, \mathcal{A}_{s}$, we have that  $\epsilon_0$ and $\lambda$ are invariant with $j.$ On the other hand, as $E(|\mathcal{A}_{j}|)=|\textsf{T}_P^n|/s=2^{n(I(X; Y)-\theta)}$, from Markov inequality,
  \begin{eqnarray}
  P(|\mathcal{A}_{j}|> 2^{n(I(X;Y)-\frac{\theta}{2})})\le 2^{-n\theta/2}.
  \end{eqnarray}

Define Boolean function $F(\mathcal{A}_{j})=1$ if and only if either $\mathcal{A}_{j}$ violates Eq. (\ref{eq: Ajgoo}) or $|\mathcal{A}_{j}|> 2^{n(I(X;Y)-\frac{\theta}{2})}.$  In other words, $F(\mathcal{A}_{j})=1$ if and only if $e(\mathcal{A}_{j})> 2^{-n\lambda \epsilon^2}$. Then, $P(F(\mathcal{A}_{j})=1)<2^{-n\theta/4+1}.$ Thus,
   \begin{eqnarray}
E\Big{(}\frac{1}{s}\sum_{j=1}^{s}F(\mathcal{A}_{j})\Big{)} \le 2^{-n\theta/4+1}.
\end{eqnarray}

Thus, by Markov inequality,
   \begin{eqnarray}
P\Big{(}\frac{1}{s}\sum_{j=1}^{s}F(\mathcal{A}_{j})>2^{-n\theta/8}\Big{)} \le 2^{-n\theta/8+1}  \label{eq: As2}
\end{eqnarray}
That is, with probability $1-2^{-n\theta/8+1}$ (over the randomness of a partition),  $\frac{1}{s}\sum_{j=1}^{s}F(\mathcal{A}_{j})\le 2^{-n\theta/8}$. In other words,
with probability $1-2^{-n\theta/8+1}$, there are at most $2^{-n\theta/8}s$ possible $j$'s with  $e(\mathcal{A}_{j})> 2^{-n\lambda\epsilon^2}$ (i.e., $F(\mathcal{A}_{j})=1$).
\end{IEEEproof}

\subsection{Proof of Theorem 1}
\noindent {\bf Proof idea. } We first explain  the idea for properties 1 and 2.   For $0<\theta<\tau$, $0<s_1\le  2^{n[I(X;Y)-I(X;Z)-\tau]}$ and
$s_2=|\textsf{ T}_P^n| 2^{-n[I(X;Y)-\theta]}$, consider independent and uniformly random partitions    $\sigma_1:{\textsf{T}_P^n}\rightarrow \{1,\cdots,s_1\}$ and $\sigma_2: {\textsf{ T}_P^n}\rightarrow \{1,\cdots,s_2\}$ for ${\textsf{T}_P^n}$.
Then, $\sigma=(\sigma_1,\sigma_2)$ is  a random partition of size $s_1s_2$ for $\textsf{T}_P^n.$ Let ${\cal A}_{ij}=\sigma^{-1}(i, j)$. Then by Lemma \ref{pro: partition},
\begin{eqnarray}
{|{\mathcal A}_{ij}|}=\frac{|\textsf{T}_P^n|}{s_1s_2}(1+\epsilon_{ij}) \label{eq: code-p1} \\
\textsf{SD}(Z^n|\sigma(X^n);Z^n)<2^{-n\beta_1}  \label{eq: code-p2}
\end{eqnarray}
for some $\beta_1>0$ and small  $\epsilon_{ij}\ge 0$.
As ${\cal A}_{\cdot j}:={\cal A}_{1j}\cup\cdots \cup {\cal A}_{s_1j}=\sigma_2^{-1}(j)$ is a random subset of $\textsf{T}_P^n$, by Lemma \ref{le: bound},
 most of ${\cal A}_{\cdot 1}, \cdots, {\cal A}_{\cdot s_2}$  are codes with small errors. If all of ${\cal A}_{\cdot 1}, \cdots, {\cal A}_{\cdot s_2}$ are codes with small errors and $\epsilon_{ij}=0$, then properties 1-2 follows by defining  ${\cal C}_{ij}={\cal A}_{ij}$, as in this case,  property 2 is just Eq. (\ref{eq: code-p2}). For the  general case, since $\epsilon_{ij}$ is small and most of  ${\cal A}_{\cdot j}$'s are good codes, we can discard  ${\cal A}_{\cdot j}$ (that is not a good code) and define ${\cal C}_{ij}$ to be ${\cal A}_{ij}$ (where ${\cal A}_{\cdot j}$ is a good code) except cutting off a small subset of ${\cal A}_{ij}$ (to make  ${\cal C}_{ij}$ having  an equal size).  As the changes are minor, the  resulting  ${\cal C}_{\cdot j}$'s  will remain  a good code and  satisfy   property 2. The main  effort in the proof is to make the above idea  rigorous.

Then, we explain the idea for property 3. We need  a fact: \\
{\bf Fact 1. } \quad For RVs $U, V$ over ${\cal U}, {\cal V}$ and function $F: {\cal U}\times {\cal V}\rightarrow \mathbb{R}^+$ with $F(u, v)\le \alpha$ for any $u, v$ and some $\alpha>0$, let $P_{UV}(u, v)\le \beta Q_{UV}(u,v)+\delta_{uv}$ for some $\beta>0, \delta_{uv}>0$ and any distributions $P_{UV}, Q_{UV}$. Then,  $$E(F(U, V))\le \alpha \sum_{u,v}\delta_{uv}+\beta\sum_{u, v}Q_{UV}(u,v)F(u, v).$$
Now we come back to the idea for property 3. Notice that    $\hat{Y}^n$ is obtained as follows. We sample $I, J$ and sample $\hat{X}^n$ from ${\cal C}_{IJ}$ randomly and finally sends $\hat{X}^n$ through channel $W_1$. As $\hat{Y}^n$ is typical with $\hat{X}^n$, with high probability $\hat{Y}^n\in \textsf{T}_{[W_1]_\epsilon}^n({\cal C}_{IJ}).$
However, under the typicality decoding $g$ (in property 1), property 3  requires  to  bound $P(\hat{Y}^n\in \textsf{T}_{[W_1]_\epsilon}^n({\cal C}_{IJ'}))$ (denoted by $\mu$).  We can define $F(i, j, u, j')=P(\hat{Y}^n\in \textsf{T}_{[W_1]_\epsilon}^n({\cal C}_{IJ'})|IJ\hat{X}^nJ'=ijuj').$ Under this, $\mu=E(F).$ By condition (b) in property 3, $P_{J'IJ\hat{X}^n}=P_{J'|IJ}P_{IJ\hat{X}^n}$, which is further equal to $P_{J'|IJ}P_{\hat{X}^n|IJ}P_IP_J=P_{J'IJ}/r,$ where $|{\cal C}_{ij}|=r$.   By Fact 1 (with $\alpha=\beta=1, Q_{UV}=P_UP_V$) and condition (a) in property 3, $\mu\le \delta_1+\mu^*$, where $\mu^*$ is  $E(F)$ with $P_{IJ\hat{X}^nJ'}$ defined as $P_IP_{JJ'}/r.$ Similarly, by condition (c) in property 3 and Fact 1 (with $\beta=2^{n^\omega}, \alpha=1$ and $Q_{JJ'}=\frac{1}{\mathbb{J}(\mathbb{J}-1)}$), we have $\mu\le \delta_1+\delta_2+2^{n^\omega}\mu'$, where $\mu'=E(F)$ with $P_{IJ\hat{X}^nJ'}=\frac{1}{r\mathbb{J}(\mathbb{J}-1)\mathbb{I}}.$  In this case, $P_{IJ\hat{X}^nJ'}$ is now explicit and  simple. We calculate based on
 Lemma \ref{le: intersec} to   show that $\mu'$ is  roughly
$r2^{-nI(X;Y)},$  which is of $2^{-n\eta}$ for some $\eta>0$, as  $I(X;Y)-I(X;Z)>0$ and we can set $r=2^{n(I(X;Z)-\eta)}$. Since $n\eta>n^\omega$ for $\omega<1$, $\mu$ is dominated by $\delta_1+\delta_2$. This completes property 3.

With the above ideas in mind, we now implement the proof details rigorously.

 \vspace{.10in} \noindent{\bf Proof. } \quad \underline{\em Part I (for properties 1-2).} From our assumption, $P_X=P$. Hence,  $P_{XY}(x, y)=P(x)W_1(y|x)$ and $P_{XZ}(x, z)=P(x)W_2(z|x)$.
  Let $\widetilde{P}_{X^nZ^n}(x^n,z^n)\stackrel{def}{=}P_{XZ}^n(x^n,z^n)/P^n_X(\textsf{T}_P^n)$ for $x^n\in \textsf{T}_P^n$ and $z^n\in \mathcal{Z}^n$. Then its   marginal distribution $\widetilde{P}_{X^n}(x^n)$ is $\widetilde{P}_{X^n}(x^n)=\frac{1}{|{\textsf{T}_P^n}|}$ {for} $x^n\in{\textsf{T}_P^n}.$

For any $\theta\in (0, \tau)$, let  $s_1, s_2$ be any integers with \begin{eqnarray*}
1\le&s_1&\le 2^{n[I(X;Y)-I(X;Z)-\tau]},\\
&s_2&=|\textsf{ T}_P^n| 2^{-n[I(X;Y)-\theta]}.
\end{eqnarray*}
 Consider independent and uniformly random partitions   of ${\textsf{T}_P^n}$, $\sigma_1:{\textsf{T}_P^n}\rightarrow \{1,\cdots,s_1\}$ and $\sigma_2: {\textsf{ T}_P^n}\rightarrow \{1,\cdots,s_2\}$.

Then, $\sigma=(\sigma_1,\sigma_2)$ is  a partition of size $s_1s_2$ for $\textsf{T}_P^n.$ Let $\mathcal{A}=\textsf{T}_P^n$.
By Lemma  \ref{pro: partition} with $Z$ in the role of $Y$ and  $\sigma=(\sigma_1,\sigma_2)$ (hence $\mathcal{B}=[s_1]\times [s_2]$ in the remark after this lemma and notice that $k=s_1s_2\le |\textsf{T}_P^n|2^{-n(I(X; Z)+(\tau-\theta))}$), there exists $n_1>0, \alpha_1>0$ and  $\beta_1>{0}$ such that the following holds with  probability  $1-2^{-n\alpha_1}$ (over $\sigma$),
\begin{eqnarray}
{|{\mathcal A}_{ij}|}=\frac{|\textsf{T}_P^n|}{s_1s_2}(1+\epsilon_{ij}) \label{eq7}\\
\textsf{SD}(Z^n|\sigma(X^n);Z^n)<2^{-n\beta_1} \label{eq8}
\end{eqnarray}
for $n\geq{n_1}$, where  $\mathcal{A}_{ij}=\sigma_1^{-1}(i)\cap\sigma_2^{-1}(j)$ and $|\epsilon_{ij}|\le 2^{-n\beta_1}$.

Let $\mathcal{A}_{\cdot j}=\cup_i \mathcal{A}_{ij}.$ Then, $\mathcal{A}_{\cdot j}=\sigma_2^{-1}(j)$ and hence $\{\mathcal{A}_{\cdot j}\}_{j=1}^{s_2}$ is the explicit representation of partition $\sigma_2$. By Lemma \ref{le: bound}, there exist  constants $\lambda>0$ and  $\epsilon_0>0$ such that
with probability $1-2^{-n\theta/8+1}$ (over $\sigma$), there are at most $2^{-n\theta/8}s_2$ possible $j$'s with  $e(\mathcal{A}_{\cdot j})> 2^{-n\lambda\epsilon^2}$,  for  any $\epsilon<\epsilon_0$.

Define $\textsf{Bad}(\sigma)$ to the event:  under  $\sigma$,  either Eqs. (\ref{eq7})(\ref{eq8}) fails,  or $e(\mathcal{A}_{\cdot j})>2^{-n\lambda\epsilon^2}$ occurs to  more than $s_2 2^{-n\theta/8}$ possible $j$'s. Then  $\Pr[\textsf{Bad}(\sigma)]\le 2^{-nc+2}$ for  $c=\min(\alpha_1, \theta/8).$

From Eqs. (\ref{eq: sd})({\ref{eq8}) and $\mathcal{B}=[s_1]\times [s_2]$, noticing
$$\widetilde{P}_{X^n}\Big{(}\sigma^{-1}_1(i)\cap \sigma^{-1}_2(j)\Big{)}=\widetilde{P}_{X^n}\Big{(}\sigma^{-1}_1(i)\Big{)}\cdot \widetilde{P}_{X^n}\Big{(}\sigma^{-1}_2(j)\Big{)}$$ (as $\sigma_1, \sigma_2$ are independent),  we have
\begin{eqnarray}
\nonumber
&&\textsf{SD}(Z^n|\sigma(X^n);Z^n)\\
\nonumber
&=&\sum_{i, j}\widetilde{P}_{X^n}(\mathcal{A}_{ij}) \textsf{SD}(\widetilde{P}_{Z^n|(i, j)};\widetilde{P}_{Z^n})\\
\nonumber
&=& \sum_{j=1}^{s_2}\widetilde{P}_{X^n}(\mathcal{A}_{\cdot j})\left(\sum_{i=1}^{s_1}\widetilde{P}_{X^n}(\mathcal{A}_{i\cdot }) \textsf{SD}(\widetilde{P}_{Z^n|(i, j)};\widetilde{P}_{Z^n})\right)\\
&\le&  2^{-n\beta_1}. \label{eq: trans}
 \end{eqnarray}
 where $\widetilde{P}_{X^n} (\mathcal{A}_{i j})=\frac{|\mathcal{A}_{ij}|}{|\textsf{T}_P^n|}=\frac{1}{s_1s_2}+\frac{\epsilon_{ij}}{s_1s_2}$ for  $|\epsilon_{ij}|\le 2^{-n\beta_1}$. Let $\bar{\epsilon}_{\cdot j}=\sum_{i=1}^{s_1} \epsilon_{ij}/s_1.$ We have  $\widetilde{P}_{X^n}(\mathcal{A}_{\cdot j})=\frac{1}{s_2}+\frac{\bar{\epsilon}_{\cdot j}}{s_2}$ with $|\bar{\epsilon}_{\cdot j}|\le 2^{-n\beta_1}.$
 Thus, as $\textsf{SD}(Q_1; Q_2)\le 2$ for any distribution $Q_1, Q_2$, Eq. (\ref{eq: trans}) implies
 \begin{eqnarray}
 \frac{1}{s_2}\sum_{j=1}^{s_2} \left(\sum_{i=1}^{s_1}\widetilde{P}_{X^n}(\mathcal{A}_{i\cdot }) \textsf{SD}(\widetilde{P}_{Z^n|(i, j)};\widetilde{P}_{Z^n})\right)< 2^{-n\beta_1+2}.  \label{eq9}
\end{eqnarray}
Similarly, we obtain   \begin{eqnarray}
 \frac{1}{s_1s_2}\sum_{j=1}^{s_2} \left(\sum_{i=1}^{s_1} \textsf{SD}(\widetilde{P}_{Z^n|(i, j)};\widetilde{P}_{Z^n})\right)< 2^{-n\beta_1+4}.  \label{eq9'}
\end{eqnarray}

 When Eq. (\ref{eq9'}) holds, Markov inequality implies  the number  of $j$'s with
\begin{eqnarray}
\frac{1}{s_1}\sum_{i=1}^{s_1}\textsf{SD}(\widetilde{P}_{Z^n|(i, j)};\widetilde{P}_{Z^n})< 2^{-n\beta_1/2+4} \label{eq: reducj}
\end{eqnarray}
is at least $s_2(1-2^{-n\beta_1/2}).$

For any $\sigma$ with $\neg\textsf{Bad}$, we already know that Eq. (\ref{eq9'}) holds and the number of $j$'s with $e(\mathcal{A}_{\cdot j})>2^{-n\lambda\epsilon^2}$ is bounded by $s_22^{-n\theta/8}$. Hence, if we let $\mathcal{J}'$ be the set of $j$ such that Eq. (\ref{eq: reducj}) holds and $e(\mathcal{A}_{\cdot j})\le 2^{-n\lambda\epsilon^2}$, then  for any $\sigma$ with $\neg\textsf{Bad}$,
\begin{eqnarray*}
|\mathcal{J}'|\ge s_2(1-2^{-n\theta/8}-2^{-n\beta_1/2}).
\end{eqnarray*}

For each $j\in \mathcal{J}'$, make $|\mathcal{A}_{ij}|=\frac{|\textsf{T}_P^n|}{s_1s_2}(1-2^{-n\beta_1})$ by cutting  a {\em uniformly random} subset of a proper size  from   $\mathcal{A}_{ij}$. After this, for $j\in \mathcal{J}'$ and $i\in [s_1]$,  denote  $\mathcal{A}_{ij}, \mathcal{A}_{i\cdot}, \mathcal{A}_{\cdot j}$ respectively by  $\mathcal{C}_{ij}, \mathcal{C}_{i\cdot}, \mathcal{C}_{\cdot j}$. Let  $\mathcal{C}=\cup_{i\in [s_1], j\in \mathcal{J}'}\mathcal{C}_{ij}$.  Also update $\tilde{P}_{X^nZ^n}(x^n, z^n)= P_{XZ}^n(x^n, z^n)/P^n_X(\mathcal{A})$ to $\tilde{P}_{X^nZ^n}(x^n, z^n)= P_{XZ}^n(x^n, z^n)/P^n_X(\mathcal{C}).$ Correspondingly update $\tilde{P}_{X^n}(x^n), \tilde{P}_{Z^n}$. Then, we have $\widetilde{P}_{X^n}(\mathcal{C}_{i\cdot})=1/s_1$. Note now $\textsf{SD}(\widetilde{P}_{Z^n|(i, j)};\widetilde{P}_{Z^n})$ is updated by a multiplicative  factor $P_X^n(\mathcal{A})/P_X^n(\mathcal{C})$. Hence,  Eq. (\ref{eq: reducj}) is now updated to
\begin{eqnarray}
\frac{1}{s_1}\sum_{i=1}^{s_1}\textsf{SD}(\widetilde{P}_{Z^n|(i, j)};\widetilde{P}_{Z^n})<2^{-n\beta_1+5}
\end{eqnarray}
 for every $j\in \mathcal{J}'$. Let $\mathcal{J}$ be a {\em uniformly random} subset of $\mathcal{J}'$ of size ${\mathbb J}$ and let ${\mathbb I}=s_1$.   Then, with probability at least $1-2^{-nc+2}$ over $\sigma$ (i.e., when $\neg\textsf{Bad}$ occurs), we get $\mathcal{J}$ s.t.
\begin{itemize}
\item[1.] For any $j\in \mathcal{J}$, $\mathcal{C}_{\cdot j}$ is a code $(f_j, g_j)$ with average error probability at most $2^{-n\lambda\epsilon^2+1}$ as the cutting treatment  on $\mathcal{A}_{ij}$ can increase the average error probability by at most $\frac{1+2^{-n\beta_1}}{1-2^{-n\beta_1}}<2.$
\item[2.] For any $j\in \mathcal{J}$,  $\frac{1}{s_1}\sum_{i=1}^{s_1}\textsf{SD}(\widetilde{P}_{Z^n|(i, j)};\widetilde{P}_{Z^n})< 2^{-n\beta_1+5}$. So, for any $P_{IJ}=P_J/s_1,$  $\textsf{SD}(\widetilde{P}_{Z^n|(J, I)};\widetilde{P}_{Z^n})< 2^{-n\beta} $ for $\beta<\beta_1$ (not depending on $P_{J})$ and $n$ large enough.
\end{itemize}
Note that  $\lim_{n\rightarrow \infty}\frac{1}{n}\log (s_2(1-2^{-n\theta/8}-2^{-n\beta_1/2}))= H(X|Y)+\theta$ and $\theta$ is arbitrary in $(0, \tau)$. So  we can define  ${\mathbb J}$ to be  any value as long as $\frac{1}{n}\log {\mathbb J}<H(X|Y)+\tau.$  So ${\mathbb J}$ and ${\mathbb I}$ can take any value in the required condition.

So far we have proved that for  $1-2^{-nc+2}$ fraction of $\sigma$ (denoted by set \textsf{Good}),  uniformly random $\mathcal{J}$ from $\mathcal{J}'$ and  uniformly random  $\mathcal{C}_{ij}$ from $\mathcal{A}_{ij}$ will satisfy  properties 1-2. Note the uniformity of $\mathcal{J}$ and $\mathcal{C}_{ij}$ is unnecessary for property 1-2 and it is for the proof of property 3 in the following.

\vspace{.10in} \noindent\underline{\em Part II (continue for property 3).}
 We continue to prove property 3,  based on set \textsf{Good}, the uniformity of  $\mathcal{C}_{ij}, \mathcal{J}$ above and properties 1-2.
  We will show that for a large fraction of \textsf{Good}, there exists some choice of $\mathcal{J}$ and $\mathcal{C}_{ij}$ (in properties 1-2) that further  satisfies  property 3.

Let $r=\frac{|\textsf{T}_P^n|}{s_1s_2}(1-2^{-n\beta_1}),$
 $\mathcal{C}_{ij}=\{u_1, \cdots, u_{r}\}$ and $\mathcal{C}_{ij'}=\{v_1, \cdots, v_{r}\}$ where elements are ordered uniformly randomly.
Since $g_j$ uses typicality decoding (Lemma \ref{le: pack}), for any $\sigma$,
\begin{eqnarray*}
&&P\Big{(}g_{J'}(\hat{Y}^n)\in \mathcal{C}_{IJ'}\Big{)}\le P\Big{(}\hat{Y}^n\in \textsf{T}_{[W]_\epsilon}(\mathcal{C}_{IJ'})\Big{)}\\
&=&\sum_{i, j', j, t}{P_{IJJ'\hat{X}^n}(i,j,j', u_t)}\times \\
&& P\Big{(}\hat{Y}^n\in \textsf{T}_{[W]_\epsilon}(\mathcal{C}_{ij'})|IJJ'\hat{X}^n=ijj'u_t\Big{)}\\
&=&\sum_{i, j', j, t}{P_{IJJ'}(i,j, j')}\frac{1}{r}\times \\
&& P\Big{(}\hat{Y}^n\in \textsf{T}_{[W]_\epsilon}(\mathcal{C}_{ij'})|\hat{X}^n=u_t\Big{)}\\
&& \mbox{($J'\rightarrow IJ\rightarrow \hat{X}^n\rightarrow \hat{Y}^n$ Markovity assumption)}\\
&\le& \sum_{i, j', j, t}\frac{P_{JJ'}(j, j') P\Big{(}\hat{Y}^n\in \textsf{T}_{[W]_\epsilon}(\mathcal{C}_{ij'})|\hat{X}^n=u_t\Big{)}}{r{\mathbb I}}\\
&& +\delta_1, \quad (\mbox{from condition (a) in property 3})
\end{eqnarray*}
Further by condition (c) in property 3,  we have
\begin{eqnarray}
\nonumber
&& P\Big{(}\hat{Y}^n\in \textsf{T}_{[W]_\epsilon}(\mathcal{C}_{IJ'})\Big{)}-\delta_1-\delta_2\\
 &\le& \sum_{i, j', j, t}\frac{2^{n^\omega} P\Big{(}\hat{Y}^n\in \textsf{T}_{[W]_\epsilon}(\mathcal{C}_{ij'})|\hat{X}^n=u_t\Big{)}}{r{\mathbb J}({\mathbb J}-1){\mathbb I}}. \label{eq: o(n)}
\end{eqnarray}

Notice $\sum_{i, j', j, t}\frac{1}{{\mathbb I}{\mathbb J}({\mathbb J}-1)r}P\Big{(}\tilde{Y}^n\in \textsf{T}_{[W]_\epsilon}(\mathcal{C}_{ij'})|\hat{X}^n=u_t\Big{)}$ equals  $P\Big{(}\hat{Y}^n\in \textsf{T}_{[W]_\epsilon}(\mathcal{C}_{IJ'})\Big{)}$ but with $P_{IJJ'\hat{X}^n}=\frac{1}{r{\mathbb I}{\mathbb J}({\mathbb J}-1)}$ (i.e., $I, (J, J')$ independent and each uniformly random and $\hat{X}^n$ uniformly random in $\mathcal{C}_{IJ}$).  We now bound $P\Big{(}\hat{Y}^n\in \textsf{T}_{[W]_\epsilon}(\mathcal{C}_{IJ'})\Big{)}$ under this setting. By Lemma \ref{le: basicH} (4),
\begin{eqnarray}
\nonumber
&&P\Big{(}\hat{Y}^n\in \textsf{T}_{[W]_\epsilon}(\mathcal{C}_{IJ'})\Big{)}-2^{-n\lambda_1\epsilon^2}\\
\nonumber
&\le& P\Big{(}\hat{Y}^n\in \textsf{T}_{[W]_\epsilon}(\hat{X}^n)\cap \textsf{T}_{[W]_\epsilon}(\mathcal{C}_{IJ'})\Big{)}\\
\nonumber
&=& \sum_{i, j', j, t}\frac{P\Big{(}\hat{Y}^n\in \textsf{T}_{[W]_\epsilon}(u_t)\cap \textsf{T}_{[W]_\epsilon}(\mathcal{C}_{ij'})|\hat{X}^n=u_t\Big{)}}{{\mathbb I}{\mathbb J}({\mathbb J}-1)r}\\
&\le & \sum_{i, j', j, t, t'}\frac{\Big{|}\textsf{T}_{[W]_\epsilon}(u_t)\cap \textsf{T}_{[W]_\epsilon}(v_{t'})\Big{|}}{{\mathbb I}{\mathbb J}({\mathbb J}-1)r2^{n(H(Y|X)-\epsilon)}}, \label{le: o(n)'}
\end{eqnarray}
 for some $\lambda_1>0$, where $\mathcal{C}_{ij}=\{u_1, \cdots, u_{r}\}$  and $\mathcal{C}_{ij'}=\{v_1, \cdots, v_{r}\}$.

Let $\xi$ be  the randomness to select $\mathcal{J}$ from $\mathcal{J}'$ and to select $\mathcal{C}_{sd}$ from $\mathcal{A}_{sd}$ for all $s, d$. Let $\eta$ be the randomness to order  elements in $\mathcal{C}_{sd}$ for all $s, d$.   So far we have assumed $\xi, \eta$ and $\sigma$ are fixed. As $J\ne J'$ (so    $u_t\ne v_{t'}$), it is not hard to see that, over the randomness of  $(\xi, \eta, \sigma)$, RV  $(u_t, v_{t'})$ for fixed $(t, t')$ has a probability distance $2^{-n\gamma}$ from a uniformly random pair $(U, V)$ in $\textsf{T}_P^n$ for some  constant $\gamma>0$.  So
\begin{eqnarray}
\nonumber
&& E\Big{(}\sum_{i, j', j, t, t'}\frac{\Big{|}\textsf{T}_{[W]_\epsilon}(u_t)\cap \textsf{T}_{[W]_\epsilon}(v_{t'})\Big{|}}{{\mathbb I}{\mathbb J}({\mathbb J}-1)r2^{n(H(Y|X)-\epsilon)}}\Big{)}\\
\nonumber
&\le & 2^{-n(\gamma-2\epsilon)}+E\Big{(}\sum_{i, j, t, j', t'}\frac{\Big{|}\textsf{T}_{[W]_\epsilon}(U)\cap \textsf{T}_{[W]_\epsilon}(V)\Big{|}}{{\mathbb I}{\mathbb J}({\mathbb J}-1)r2^{n(H(Y|X)-\epsilon)}}\Big{)}\\
\nonumber
&\le& 2^{-n\gamma/2}+r2^{-n(I(X; Y)-c'\epsilon)}, \quad(\mbox{$c'$ constant, Lemma \ref{le: intersec}}) \\
\nonumber
&=& 2^{-n\gamma/2}+2^{-n(\theta-c'\epsilon)}\le 2^{-n\gamma''+1},
\end{eqnarray}
for $\gamma''<\min\{\gamma/2, \theta/2\}.$   So for  $1-2^{-n\gamma''/2+1}$ fraction of $\sigma$, there exists $\xi$ and $\eta$ so that
 \begin{equation}
\sum_{i, j', j, t, t'}\frac{{|}\textsf{T}_{[W]_\epsilon}(u_t)\cap \textsf{T}_{[W]_\epsilon}(v_{t'}){|}}{{\mathbb I}{\mathbb J}({\mathbb J}-1)r2^{n(H(Y|X)-\epsilon)}}\le 2^{-n\gamma''/2}. \label{eq: o(n)''}
\end{equation}
 Denote this set of $\sigma$ by \textsf{Good}$'$.  Then for  $\sigma\in \textsf{Good}\cap \textsf{Good}'$,  from Eq. (\ref{le: o(n)'})(\ref{eq: o(n)''}), we know that  Eq. (\ref{eq: o(n)}) is bounded by $2^{-n\lambda\epsilon^2+n^\omega}+2^{-n\gamma''/2+n^\omega}.$ Hence, property 3 is satisfied if we take $\epsilon=\sqrt[3]{\frac{1}{n^{1-\omega}}}$,  as $2^{n^\omega-n\gamma''/4}+2^{-n\lambda_1\epsilon^2+n^\omega}<2^{-n^\omega}$ when  $n$ large enough.

As a summary, for $P(\textsf{Good}\cap \textsf{Good}')>1-2^{-nc+2}-2^{-n\gamma''/2+1}$ fraction of $\sigma$,  properties 1-3 are satisfied.    $\hfill{\blacksquare}$


\end{document}